\documentclass[nofootinbib, pra]{revtex4} 
\usepackage{amssymb, amsmath}
\usepackage{graphicx}
\usepackage{verbatim}
\usepackage{xcolor}
\begin{document}
\title{Physical interpretation of non-normalizable harmonic oscillator states and relaxation to pilot-wave equilibrium}
\author{Indrajit Sen}
\email{isen@chapman.edu}
\affiliation{Institute for Quantum Studies, Chapman University\\
One University Drive, Orange, CA, 92866, USA}
\date{\today}
\begin{abstract}
Non-normalizable states are difficult to interpret in the orthodox quantum formalism but often occur as solutions to physical constraints in quantum gravity. We argue that pilot-wave theory gives a straightforward physical interpretation of non-normalizable quantum states, as the theory requires only a normalized density of configurations to generate statistical predictions. In order to better understand such states, we conduct the first study of non-normalizable solutions of the harmonic oscillator from a pilot-wave perspective. We show that, contrary to intuitions from orthodox quantum mechanics, the non-normalizable eigenstates and their superpositions are bound states in the sense that the velocity field $v_y \to 0$ at large $\pm y$. We argue that defining a physically meaningful equilibrium density for such states requires a new notion of equilibrium, named pilot-wave equilibrium, which is a generalisation of the notion of quantum equilibrium. We define a new $H$-function $H_{pw}$, and prove that a density in pilot-wave equilibrium minimises $H_{pw}$, is equivariant, and remains in equilibrium with time. We prove an $H$-theorem for the coarse-grained $H_{pw}$, under assumptions similar to those for relaxation to quantum equilibrium. We give an explanation of the emergence of quantization in pilot-wave theory in terms of instability of non-normalizable states due to perturbations and environmental interactions. Lastly, we discuss applications in quantum field theory and quantum gravity, and implications for pilot-wave theory and quantum foundations in general.
\end{abstract}

\flushbottom
\maketitle

\thispagestyle{empty}

\section{Introduction}

\label{intro}
Pilot-wave theory (also called de Broglie-Bohm theory or Bohmian mechanics) is a realist, nonlocal formulation of quantum mechanics originally presented in the 1927 Solvay conference by de Brogile \cite{bell, solventini}. In 1952, Bohm showed how the theory solves the vexed measurement problem in orthodox quantum mechanics by describing the measurement apparatus within the theory \cite{bohm1, bohm2}. The theory has been extended to the relativistic domain \cite{bohm53, bohmbook2, valentiniphd, struyvefields, durr14}, applied to astrophysical and cosmological scenarios \cite{valentiniastro, pintu19, pk20, teenv}, and provides a counter-example to the claim that quantum phenomena imply a denial of realism. \\

In his description of the theory, Bohm pointed out that certain assumptions are necessary to reproduce orthodox quantum mechanics. Further, he opined that these assumptions may need modifications in regimes not yet experimentally accessible, so that the theory may either supersede or depart from orthodox quantum mechanics in the future \cite{bohm1, bohm2, bohm53, bohmrelax}. One of these assumptions is that the initial density of configurations equals the Born rule density. This assumption has been criticised on the grounds that, since there is no logical relation between the initial configuration density and the quantum state in the theory, it is ad hoc \cite{killer, pauli53}. Bohm was able to show that adding random collisions \cite{bohmrelax} or random fluid fluctuations \cite{bohm54} to the dynamics of the theory leads to relaxation from an arbitrary density to the Born rule density. Later, Valentini showed that the original dynamics alone is sufficient for relaxation to occur at a coarse grained level \cite{valentinI, royalvale}. Numerous computational studies have since been conducted that have furthered our understanding of the relaxation process in various scenarios (see \cite{teenv} for a review).\\

However, a simple but important conceptual point has remained largely unnoticed in the literature: if there is no logical relationship between the configuration density and the quantum state in pilot-wave theory, then why should the quantum state be normalizable? In orthodox quantum mechanics, normalizability is necessary as statistical predictions are extracted from the quantum state according to the Born rule. On the other hand, in pilot-wave theory the quantum state serves as a physical field that determines the evolution of the configuration. To extract statistical predictions from the theory, one only needs to define an \textit{ensemble with a normalized density of configurations} -- normalizability of the quantum state is unnecessary. This opens up the possibility of physically interpreting non-normalizable quantum states that occur as solutions to physical constraints in quantum gravity, such as the Kodama state \cite{kodamaginal, witten003, allin}.\\

However, to the best of our knowledge, the behaviour of non-normalizable solutions to the Schrodinger equation has not been studied from a pilot-wave perspective. In this article, we make a first step in this direction by studying the non-normalizable solutions of the harmonic-oscillator potential. We choose the harmonic oscillator as it is widely found in nature, and because the normalizability constraint leads to the important discretization of the energy levels. The article is structured as follows. We study the non-normalizable solutions of the harmonic oscillator, using both the analytic approach and the ladder operator approach. We then study the pilot-wave theory of the non-normalizable states. We show in this section that the pilot-wave velocity field for the non-normalizable states $v_y \to 0$ at large $\pm y$. We discuss the relaxation behaviour for these states. We then introduce the notion of pilot-wave equilibrium and define the new $H$-function $H_{pw}$. We prove an $H$-theorem applicable to non-normalizable states using a coarse-grained $H_{pw}$, analogous to the $H$-theorem for quantum equilibrium. We study the relationship between relaxation to pilot-wave equilibrium and relaxation to quantum equilibrium. Lastly, we discuss the theoretical and experimental implications of our work. In particular, we show that non-normalizable states are unstable in the presence of perturbations and environmental interactions, and thereby give an explanation of quantization in pilot-wave theory.\\

\section{Non-normalizable solutions of the harmonic oscillator} \label{see}
We start by noting that several elementary theorems in orthodox quantum mechanics are no longer applicable once the normalizability constraint on quantum state is dropped. In the non-normalizable scenario, eigenstates in one dimension are generally degenerate and complex as relevant theorems on degeneracy and reality of eigenstates no longer apply. Furthermore, a non-normalizable quantum state does not have a Fourier transform, and therefore a momentum representation, in general. This is because Fourier transform exists only if the concerned function does not diverge faster than a polynomial at large values of its argument. Therefore, we are restricted to the position representation of the quantum state in general. This makes sense from a pilot-wave perspective, as the position basis is the preferred basis in the theory. We also note that the momentum operator is in general non-Hermitian in this scenario.\\

For the harmonic-oscillator potential, the energy eigenvalues are not quantized and can also take negative values in this scenario. Mathematically, the eigenvalues can also be complex in this scenario, but this is not physically meaningful from a pilot-wave perspective. Consider a von-Neumann energy measurement, which leads to apparatus wavefunctions of the form $\psi(y- gEt, 0)$, where $E$ is the energy eigenvalue and $g$ is the strength of interaction between the system and apparatus. The wavefunction $\psi(y- gEt, 0)$ is not defined only on configuration space if $E$ is complex. Therefore, allowing complex eigenvalues is only possible if one abandons the configuration space as the fundamental arena of pilot-wave theory. Lastly, we restrict the initial wavefunction to only eigenstates and finite superpositions, as the time-evolution operator $e^{-i\hat{H}t/\hbar}$ may not be well-defined for any arbitrary initial wavefunction \cite{frizzy}. With these facts in mind, let us study the non-normalizable solutions to the harmonic oscillator from a pilot-wave perspective. \\

The time-independent Schrodinger equation for the harmonic-oscillator potential can be written as
\begin{align}
    -\frac{d^2\psi}{dy^2} + y^2\psi = K\psi \label{ho}
\end{align}
where $y\equiv \sqrt{m\omega/\hbar}x$ and $K \equiv 2E/\hbar\omega$. The equation is traditionally solved by using the ansatz $e^{-y^2/2}h^K(y)$. Substituting the ansatz into equation (\ref{ho}), we get
\begin{align}
    \frac{d^2 h^K}{dy^2} -2y\frac{dh^K}{dy} + (K-1)h^K = 0 \label{her}
\end{align}
Equation (\ref{her}) is known as the Hermite differential equation. It contains both normalizable and non-normalizable solutions to (\ref{ho}). Using the Frobenius method, the general solution to (\ref{her}) can be written as
\begin{align}
     h^K(y) &= \sum_{n=0}^{\infty} a_n y^n \label{solb}
 \end{align}    
 where $a_0$ and $a_1$ are two arbitrary complex constants and the recurrence relation between $a_n$'s can be obtained to be $a_{n+2} = (2n+1-K)a_n/(n+1)(n+2)$. It is useful for us to rewrite equation (\ref{solb}) as 
 \begin{align}
    h^K(y) &= a_0 \sum_{n=0}^{\infty} \frac{a_{2n}}{a_0} y^{2n} + a_1 \sum_{n=0}^{\infty} \frac{a_{2n+1}}{a_1} y^{2n+1}\\
    &= a_0 (1+ \sum_{n=1}^{\infty} \frac{\prod_{j=0}^{n-1} (4j+1-K)}{(2n)!} y^{2n}) + a_1 (y + \sum_{n=1}^{\infty} \frac{\prod_{j=0}^{n-1} (4j+3-K)}{(2n+1)!} y^{2n+1}) \label{use}\\
    &= a_0 h_0^K(y)  + a_1 h_1^K(y)  \label{sol}
 \end{align}
where $h_0^K = (1+ \sum_{n=1}^{\infty} \frac{\prod_{j=0}^{n-1} (4j+1-K)}{(2n)!} y^{2n})$ and $h_1^K = (y + \sum_{n=1}^{\infty} \frac{\prod_{j=0}^{n-1} (4j+3-K)}{(2n+1)!} y^{2n+1})$. Clearly, the term $h_0^K$ consists only of even powers of $y$, whereas the term $h_1^K$ consists only of odd powers.\\

It is useful to note that $h_0^K(y)$, $h_1^K(y)$ can be expressed in closed form as follows:
\begin{align}
h_0^K(y) &= M(\frac{1}{4}(1-K), \frac{1}{2}, y^2)\\
h_1^K(y) &= y M(\frac{1}{4}(3-K), \frac{3}{2}, y^2)
\end{align}
where 
\begin{align}
M(c, d, y) \equiv \sum_{j=0}^{\infty} \frac{(c)_j}{(d)_j} \frac{y^j}{j!}
\end{align}
is the confluent hypergeometric function of the first kind and $(t)_j \equiv \Gamma(t+j)/\Gamma(t)$ is the Pochhammer symbol.\\

The general solution to the time-independent Schrodinger equation (\ref{ho}) can be written as 
\begin{align}
     \psi^K(y) = a_0 \varphi_0^K(y)  + a_1 \varphi_1^K(y) \label{solg}
\end{align}
where $\varphi_0^K \equiv e^{-y^2/2} h_0^K(y)$ and $\varphi_1^K \equiv e^{-y^2/2} h_1^K(y)$. Equation (\ref{solg}) is a valid solution to the Schrodinger equation (\ref{ho}) for all (real) values of $K$. It can be shown that the series $h_0^K(y)$ ($h_1^K(y)$) terminates only if $K = (2n+1)$ for an even (odd) $n$. In that case, $\varphi_0^K(y)$ ($\varphi_1^K(y)$) has a $e^{-y^2/2}$ dependence at large $\pm y$ and is normalizable. If $K \neq (2n+1)$ for an even (odd) $n$, then $\varphi_0^K(y)$ ($\varphi_1^K(y)$) has a $e^{y^2/2}$ dependence at large $\pm y$ and is non-normalizable.\\

The complex coefficients $a_0$, $a_1$ contain a total of 4 real parameters. We can eliminate 2 of the parameters by \textit{a)} normalizing the coefficients so that $|a_0|^2 + |a_1|^2 = 1$ (note that the quantum state is itself non normalizable in general) and \textit{b)} eliminating the global phase. Both steps $a)$ and $b)$ make sense from a pilot-wave theory perspective as the pilot-wave velocity field $v(y) = j(y)/|\psi(y)|^2$, where $j(y)$ is the quantum probability current (see equation (\ref{vel}) below), does not depend on the global magnitude or the global phase of the quantum state. That is, a transformation of the form $\psi(y) \to \alpha \psi(y)$, where $\alpha$ is a complex constant, does not change $v(y)$. Therefore, we may further simplify equation (\ref{solg}) to
\begin{align}
     \psi^K_{\theta, \phi}(y) = \cos{\theta} \varphi_0^K(y)  + \sin{\theta} e^{i\phi} \varphi_1^K(y)  \label{sols}
\end{align}
where $\cos{\theta} = |a_0|/\sqrt{|a_0|^2 +|a_1|^2}$, $\sin{\theta} = |a_1|/\sqrt{|a_0|^2 +|a_1|^2}$, $\phi = -i\ln(a_1 |a_0|/a_0|a_1|)$ and $\theta \in [0,\pi]$, $\phi \in [0,2\pi)$. In this form, it is clear that $\varphi_0^K(y)$ and $\varphi_1^K(y)$ act as basis vectors of the doubly degenerate subspace corresponding to $K$. We note that, in orthodox quantum mechanics, steps \textit{a)} and \textit{b)} are justified (for normalizable states) on the grounds that $|\psi(y)|^2$ is a probability density. Clearly, $|\psi(y)|^2$ cannot be interpreted as a probability density in our case but \textit{a)}, \textit{b)} are still valid from a pilot-wave perspective. \\

We can connect the general solution (\ref{sols}) to the allowed solutions in orthodox quantum mechanics as follows. We know that the allowed energy levels in orthodox quantum mechanics are given by $K(n) = (2n + 1)$, where $n$ is a non-negative integer. Furthermore, we know from the preceding discussion that for all even $n$, $\varphi_0^{K(n)}(y)$ is normalizable and $\varphi_1^{K(n)}(y)$ is non-normalizable. Similarly, for odd $n$, $\varphi_1^{K(n)}(y)$ is normalizable and $\varphi_0^{K(n)}(y)$ is non-normalizable. Therefore,
\begin{align}
    \Psi_n(y) = \begin{cases} \label{def}
    N_n \varphi_0^{K(n)} & \text{, if $n$ even}\\
    N_n \varphi_1^{K(n)} & \text{, if $n$ odd}
    \end{cases}
\end{align}
where $\Psi_n(y)$ is the $n^{th}$ harmonic-oscillator eigenstate in orthodox quantum mechanics, and $N_n$ is the relevant normalization constant.\\
 
Let us consider a superposition of eigenstates corresponding to different values of $K$. Suppose $\psi(y) = \sum_n c_n \psi^{K_n}_{\theta_n, \phi_n}(y)$. As before, we normalize the coefficients ($\sum_n |c_n|^2 = 1$) and eliminate the global phase of $\psi(y)$, as the velocity field is unaffected by these changes. We also know, from the time-dependent Schrodinger equation, that $\psi(y)$ will evolve as
\begin{align}
     \psi(y,t) = \sum_n c_n e^{-i K_n\omega t/2} \psi^{K_n}_{\theta_n, \phi_n}(y)
\end{align}
 
Lastly, it is straightforward to extend the discussion to a system of $N$ particles, each in a harmonic oscillator potential. Consider the quantum state
\begin{align}
      \psi(y_1, y_2,...y_N) = \sum_m c_m \prod_{j=1}^N \psi_{\theta_j^m, \phi_j^m}^{K_j^m}(y_j) \label{np}
\end{align}
We normalize the coefficients $c_m$ and eliminate the global phase of $\psi(y_1, y_2,...y_N)$. The time evolution of $\psi(y_1, y_2,...y_N)$ can be easily calculated by the time-dependent Schrodinger equation.

\section{Bound-state interpretation of non-normalizable harmonic oscillator states} \label{pwtheory}
In pilot-wave theory, the quantum state serves to define the velocity field for the evolution of the system configuration. This can be a configuration of particles, as in pilot-wave theory of non-relativistic quantum mechanics, or a configuration of fields, as in pilot-wave theory of quantum field theory. Let us consider a system of $N$ particles in the harmonic oscillator potential with the quantum state (\ref{np}). Without loss of generality, we suppose that all the particles have the same mass $m$ for simplicity. The time-dependent Schrodinger equation implies the continuity equation
\begin{align}
\partial_t |\psi(\vec{y}, t)|^2 + \vec{\nabla}\cdot\vec{j}(\vec{y}, t) =0 \label{cont}
\end{align}
where $\vec{y} = (y_1, y_2, ... y_N)$ is a point on the configuration space, and the current 
\begin{align}
\vec{j}(\vec{y}, t) = \frac{\hbar}{2mi}\big [\bar{\psi}(\vec{y}, t)\vec{\nabla} \psi(\vec{y}, t) - \psi(\vec{y}, t)\vec{\nabla} \bar{\psi}(\vec{y}, t) \big ]
\end{align}
is defined in terms of $\vec{\nabla} = \sum_{i=1}^N \hat{y}_i \partial/\partial y_i $ and $\bar{\psi}(\vec{y}, t)$ which is the complex conjugate of $\psi(\vec{y}, t)$. From equation (\ref{cont}), the quantity
\begin{align}
\vec{v}(\vec{y}, t) \equiv \frac{\vec{j}(\vec{y}, t)}{|\psi(\vec{y}, t)|^2} \label{vel}
\end{align}
is defined as the pilot-wave velocity field. Let us consider an ensemble of the $N$-particle harmonic oscillator systems. As there is no \textit{a priori} relationship between the quantum state and the configuration density in pilot-wave theory, we can define an initial normalized density $\rho(\vec{y}, 0)$ for the ensemble. Equation (\ref{vel}) supplies the velocity field to evolve $\rho(\vec{y}, t)$: 
\begin{align}
\partial_t \rho(\vec{y}, t) + \vec{\nabla}\cdot\big (\rho(\vec{y}, t)\vec{v}(\vec{y}, t)\big ) =0
\end{align}

Clearly, experimental probabilities are well-defined as $\rho(\vec{y}, t)$ is normalized. However, there remains the question whether the velocity field (\ref{vel}) behaves physically for non-normalizable states. One example of an unphysical behaviour would be if $v_{y_i}(\vec{y}, t)$ increases with $y_i$ as $\sim y_i^{1+\epsilon}$ ($\epsilon >0$) for $i \in \{1, 2,...N\}$. In that case, the system configuration will escape to $y_i \to \infty$ in finite time. In orthodox quantum mechanics, we know that such behaviour cannot occur as the normalizability constraint ensures that the probability density $|\psi(\vec{y}, t)|^2 \to 0$ as $y_i \to \pm \infty$. For this reason, the normalizable states are referred to as \textit{bound states} in orthodox quantum mechanics.\\

We can straightforwardly generalise the definition of bound state to the non-normalizable scenario: if the velocity field (\ref{vel}) defined by $\psi(\vec{y}, t)$ is such that $v_{y_i}(\vec{y}, t) \to 0 $ in the limit $y_i \to \pm \infty$ for all $i \in \{1, 2,...N\}$, then $\psi(\vec{y}, t)$ is a bound state. Such a velocity field ensures that any initial normalized configuration density $\rho(\vec{y}, 0)$ will evolve to $\rho(\vec{y}, t)$ such that $\rho(\vec{y}, t) \to 0$ as $y_i \to \pm \infty$ for all $i \in \{1, 2,...N\}$. That is, the system configuration $\vec{y}$ remains bounded at all (finite) times.\\

Below, we prove that the non-normalizable solutions of the harmonic oscillator are bound states in this sense. We begin with the simplest case, that of an eigenstate in one dimension.

\subsection{Velocity field of an eigenstate in one-dimension}
Let us consider the velocity field of a harmonic oscillator eigenstate $\psi^K_{\theta, \phi}(y)$. We know from orthodox quantum mechanics that the normalizable eigenstates $\Psi_n(y)$ defined by (\ref{def}) are real. This implies that, for these states, the velocity field is zero everywhere and the particle is stationary. However, $\psi^K_{\theta, \phi}(y) = \cos{\theta} \varphi_0^K(y)  + \sin{\theta} e^{i\phi} \varphi_1^K(y)$ is complex in general. This implies that the velocity field for non-normalizable eigenstates is non-zero in general. Let us then calculate this velocity field.\\

We first note the general result that, if $\psi(y)$ is an eigenstate of the Hamiltonian, then
\begin{align}
\bar{\psi(y)}\psi'(y) -\psi(y)\bar{\psi}'(y) = c \text{ (constant)} \label{indet}
\end{align}
In orthodox quantum mechanics, $c=0$ as $\psi(y) \to 0$ as $y \to \infty$. In our case, on the other hand, $\psi(y) \to \infty$ as $y \to \infty$ so that the left-hand side of equation (\ref{indet}) becomes indeterminate at $y \to \infty$. However, it is convenient to evaluate the left-hand side of (\ref{indet}) for $\psi^K_{\theta, \phi}(y)$ at $y=0$. This is because the following readily verifiable calculations
\begin{align}
\varphi_0^K(0) &= 1\\
\varphi_1^K(0) &= 0\\
\frac{d\varphi_0^K(0)}{dy} &= 0\\
\frac{d\varphi_1^K(0)}{dy} &= 1
\end{align}
imply that
\begin{align}
\bar{\psi}^K_{\theta, \phi}(0)\psi'^K_{\theta, \phi}(0) -\psi^K_{\theta, \phi}(0)\bar{\psi'}^K_{\theta, \phi}(0) = 2i\cos{\theta}\sin{\theta}\sin \phi = c \label{garam}
\end{align} 
so that the current $j(y)$ is constant and independent of $K$.

\begin{figure}
\includegraphics[scale=0.5]{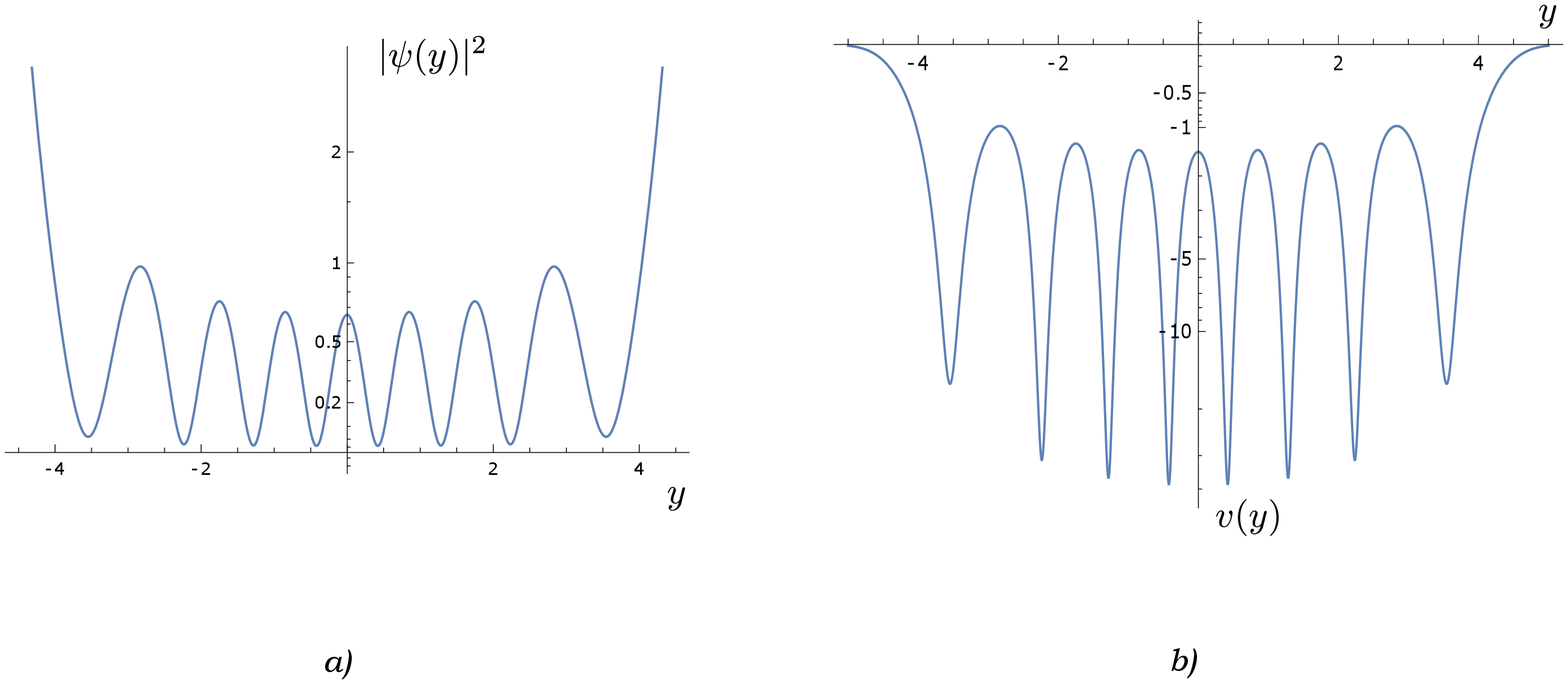}
\caption{Schematic illustration of $a)$ $|\psi(y)|^2$ and $b)$ $v(y)$ for the sample non-normalizable eigenstate $\psi(y) = \psi^{14}_{16\pi/5, 3\pi/2}(y)$. Note that $v(y) \to 0$ at large $\pm y$.}
\end{figure}

Therefore, the velocity field is 
\begin{align}
v(y, t) &= \frac{j(y)}{|\psi^K_{\theta, \phi}(y,t)|^2}\\
&= \frac{\hbar}{2mi} \frac{\bar{\psi}^K_{\theta, \phi}(y,t)\psi'^K_{\theta, \phi}(y,t) -\psi^K_{\theta, \phi}(y,t)\bar{\psi'}^K_{\theta, \phi}(y,t)}{|\psi^K_{\theta, \phi}(y,t)|^2}\\
&= \frac{\hbar}{2mi} \frac{\bar{\psi}^K_{\theta, \phi}(0,t)\psi'^K_{\theta, \phi}(0,t) -\psi^K_{\theta, \phi}(0,t)\bar{\psi'}^K_{\theta, \phi}(0,t)}{|\psi^K_{\theta, \phi}(y,t)|^2} \label{d1}\\ 
&= \frac{\hbar}{m}\frac{\cos{\theta}\sin{\theta}\sin \phi}{|\psi^K_{\theta, \phi}(y,0)|^2} \label{d2}
\end{align}
where, in equation (\ref{d2}), we have used $\psi^K_{\theta, \phi}(y,t) = e^{-iKwt/2}\psi^K_{\theta, \phi}(y,0)$ and (\ref{garam}). \\

Let us discuss the velocity field (\ref{d2}). First, equation (\ref{d2}) tells us that, for an eigenstate corresponding to $K$, the velocity field is constant with time. Second, it tells us that the velocity field depends on the angles $\theta$, $\phi$, so that degenerate eigenstates corresponding to the same $K$ will, in general, have velocity fields that are different but proportional to each other at every $y$. Third, the velocity field does not change sign with $y$. Fourth, we note that the velocity field for an eigenstate corresponding to $K = -K_0$ ($K_0 >0$) has no apparent connection with the velocity field for an eigenstate corresponding to $K = +K_0$. Lastly, and most importantly, equation (\ref{d2}) tells us that the velocity fields are inversely proportional to $|\psi^K_{\theta, \phi}(y,0)|^2$. This implies that, for $y \to \pm \infty$
\begin{align}
v(y, t) \sim \frac{\hbar}{m}\frac{\cos{\theta}\sin{\theta}\sin \phi}{e^{y^2}} 
\end{align}
as we know that $\psi^K_{\theta, \phi}(y,t)$ diverges like $\sim e^{y^2/2}$ at large $\pm y$. Therefore, the velocity field decreases very quickly to $0$ as $|\psi^K_{\theta, \phi}(y,0)|^2$ becomes large at $y \to \pm \infty$ (see Fig. 1). This implies that $\psi^K_{\theta, \phi}(y,0)$ is a \textit{bound state}, according to our definition, although it is non-normalizable. This is a surprising behaviour from the viewpoint of orthodox quantum mechanics, as a naive application of the Born rule would imply an infinitely large probability of the particle being found at large $\pm y$. 

\subsection{Velocity field of a superposition of eigenstates}
Let us consider a quantum state $\psi(y, t) = \sum_j c_j(t) \psi^{K_j}_{\theta_j, \phi_j}(y)$ that is a superposition of eigenstates corresponding to various $K$'s. We know from equation (\ref{vel}) that the velocity field is 
\begin{align}
v(y, t) = \frac{\hbar}{2mi} \frac{\bar{\psi}(y,t)\psi'(y,t) -\psi(y,t)\bar{\psi'}(y,t)}{|\psi(y,t)|^2} \label{v1}
\end{align}

To study the asymptotic behaviour of (\ref{v1}) as $y \to \pm \infty$, we first need an asymptotic expression for $\psi(y)$ as $y \to \pm \infty$. We derive such an expression in the supplementary material, using the approach given in ref. \cite{hassanahi}.

\begin{figure}
\includegraphics[scale=0.6]{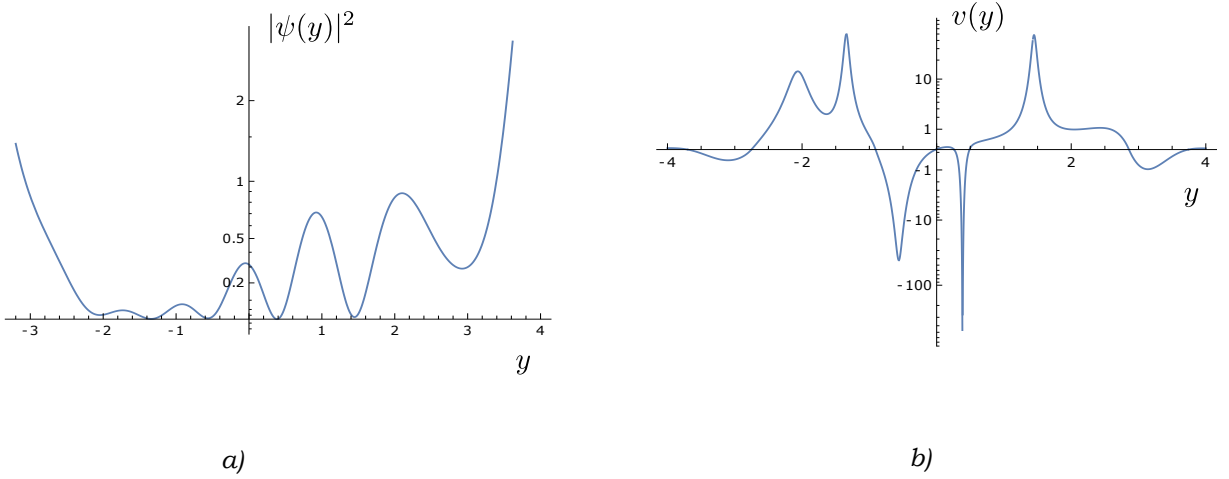}
\caption{Schematic illustration of \textit{a)} $|\psi(y)|^2$ and \textit{b)} $v(y)$ for a sample superposition $\psi(y) = 1/\sqrt{6} \psi^{15.2}_{\pi/3, \pi/4} (y) + \sqrt{2/3} e^{i\pi/5} \psi^{5.8}_{\pi/2, \pi} (y) + 1/\sqrt{6} e^{i\pi/8} \psi^{10.2}_{\pi/7, \pi/5} (y)$. Note that $v(y) \to 0$ at large $\pm y$.}
\end{figure}

\subsubsection{Asymptotic behaviour of the velocity field} \label{asymp}
Using the expansion $\psi(y,t) = \sum_j c_j(t) \psi^{K_j}(y)$, we can express the current as
\begin{align}
j(y, t) &= \frac{\hbar}{2mi}\bar{\psi}(y,t)\psi'(y,t) -\psi(y,t)\bar{\psi'}(y,t)\\
&= \frac{\hbar}{2mi} \sum_{l, j} \bar{c}_l c_j \big [\bar{\psi}^{K_l}(y)\psi'^{K_j}(y) -\psi^{K_j}(y)\bar{\psi'}^{K_l}(y) \big ] \label{cur}
\end{align}

Using the asymptotic form derived in the supplementary material, we write $\psi^{K_j}(y) \approx e^{\frac{y^2}{2}}y^{-\frac{1+K}{2}}[1 + \frac{(3+K)(1+K)}{16y^2}]$ at large $\pm y$, equation (\ref{cur}) becomes
\begin{align}
j(y, t) &\approx \frac{\hbar}{2mi} \sum_{l, j} \bar{c}_l(t) c_j(t) \frac{e^{y^2}(K_l - K_j)}{2y^2\sqrt{y^{K_j}}\sqrt{y^{K_l}}}
\end{align}
where we have retained only the leading order of $y$. Similarly, we can prove that 
\begin{align}
|\psi(y,t)|^2 \approx \sum_{l, j} \bar{c}_l(t) c_j(t)  \frac{e^{y^2}}{y\sqrt{y^{K_j}}\sqrt{y^{K_l}}}
\end{align}
Therefore, the velocity field 
\begin{align}
v(y, t)= \frac{j(y, t)}{|\psi(y,t)|^2} \sim \frac{1}{y} \text{ at large $\pm y$} \label{omfg}
\end{align}
Equation (\ref{omfg}) implies that $\lim_{y \to \pm \infty} v(y, t) = 0$ (see Fig. 2). Therefore, a superposition of eigenstates corresponding to different $K$'s is a bound state. Let us proceed next to the case of multiple particles.
\begin{figure}
\centering
\includegraphics[scale=0.5]{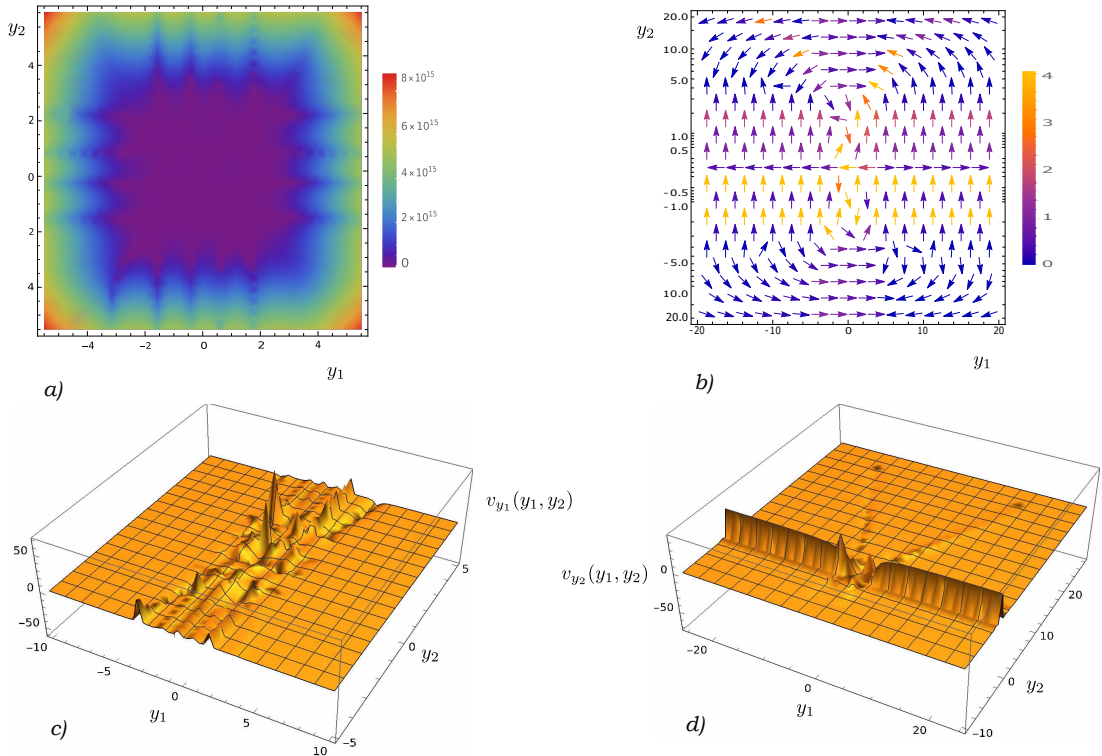}
\caption{Schematic illustration of \textit{a)} density plot for $|\psi(y_1, y_2)|^2$, \textit{b)} velocity plot for $\vec{v}(y_1, y_2)$, \textit{c)} $y_1$-velocity field $v_{y_1}(y_1, y_2)$ and \textit{d)} $y_2$-velocity field $v_{y_2}(y_1, y_2)$ for a sample superposition $\psi(y_1, y_2) = \sqrt{2}/3 \psi^{1.4}_{3\pi/4, 4\pi/3}(y_1) \psi^8_{2.2\pi, 4.1\pi}(y_2) + \sqrt{2}/3 e^{i\pi/5}\psi^{5}_{8\pi/5, 5.8\pi}(y_1) \psi^{15.6}_{2\pi/5, 9\pi/16}(y_2) + 1/3 e^{i\pi/8}\psi^{9}_{\pi/5, \pi/7}(y_1) \psi^{0.75}_{\pi/6, \pi/9}(y_2) + 2/3 e^{i\pi/9}\psi^{11.4}_{5\pi/3, 6\pi/7}(y_1) \psi^{12.6}_{2\pi/5, 7\pi/16}(y_2)$. Note, from figures \textit{b)}, \textit{c)} and \textit{d)}, that $v_{y_1}(y_1, y_2) \to 0$ at large $\pm y_1$ and $v_{y_2}(y_1, y_2) \to 0$ at large $\pm y_2$.}
\end{figure}
\subsection{Velocity field for multiple particles} \label{malty}
We want to check whether the asymptotic behaviour of the velocity field discussed in the previous subsections also hold in the case of multiple particles, each in a harmonic oscillator potential. Consider an $N$-particle quantum state
\begin{align}
\psi(\vec{y}, t) = \sum_{j=1}^n c_j(t) \prod_{g=1}^N \psi^{K_j^g}(y_g) \label{khur}
\end{align}
where $\psi^{K_j^g}(y_g)$ is an eigenstate of the $g^{th}$ particle corresponding to the eigenvalue ${K_j^g}$ in the $j^{th}$ term of the superposition. We know that the current in the $r^{th}$ direction is

\begin{align}
j_r(\vec{y}, t) &= \frac{\hbar}{2mi} \sum_{l, j} \bar{c}_l(t) c_j(t) \prod_{g \neq r}^N \psi^{K_j^g}(y_g)  \prod_{w\neq r}^N \bar{\psi}^{K_l^w}(y_w) \big [\bar{\psi}^{K_l^r}(y_r)\psi'^{K_j^r}(y_r) -\psi^{K_j^r}(y_r)\bar{\psi'}^{K_l^r}(y_r) \big ] \label{daant}
\end{align}
Similar to the previous subsection, we can express $\psi^{K_j^r}(y_r) \approx e^{y_r^2/2}y_r^{-\frac{1+(K_j^r)^2}{2}}[1 + (3+K_j^r)(1+K_j^r)/(16y_r^2)]$ at large $\pm y_r$, and then simplify (\ref{daant}) as
\begin{align}
j_r(\vec{y}, t) &\approx \frac{\hbar}{2mi} \sum_{l, j} \bar{c}_l(t) c_j(t) \prod_{g \neq r}^N \psi^{K_j^g}(y_g)  \prod_{w\neq r}^N \bar{\psi}^{K_l^w}(y_w) \frac{e^{y_r^2}(K_l^r - K_j^r)}{2y_r^2 \sqrt{y_r^{K_j^r}}\sqrt{y_r^{K_l^r}}}
\end{align}
On the other hand, 
\begin{align}
|\psi(\vec{y},t)|^2 \approx \sum_{l, j} \bar{c}_l(t) c_j(t) \prod_{g \neq r}^N \psi^{K_j^g}(y_g)  \prod_{w\neq r}^N \bar{\psi}^{K_l^w}(y_w) \frac{e^{y_r^2}}{y_r\sqrt{y_r^{K_j^r}}\sqrt{y_r^{K_l^r}}} 
\end{align}
which implies that
\begin{align}
v_r(\vec{y}, t)= \frac{j_r(\vec{y}, t)}{|\psi(\vec{y},t)|^2} \sim \frac{1}{y_r} \text{ at large $\pm y_r$} \label{vel0}
\end{align}
Equation (\ref{vel0}) confirms that the velocity field is such that $v_r(\vec{y}, t) \to 0$ as $y_r \to \infty$ $\forall r \in \{1, 2, ...N\}$. Therefore, the system configuration $\vec{y}$ remains bounded at all times and $\psi(\vec{y}, t)$ is a bound state (see Fig. 3). \\

\section{Relaxation to equilibrium} \label{relaxbe}
In pilot-wave theory for normalizable quantum states, it is well known that an arbitrary initial density of configurations relaxes to the Born rule density $|\psi(\vec{y}, t)|^2$ (called the equilibrium density) at a coarse-grained level, subject to standard statistical mechanical assumptions \cite{valentinI, royalvale, teenv}. In this section, we look at whether such a relaxation occurs to a well-defined equilibrium density when $\psi(\vec{y}, t)$ is non-normalizable. \\

\subsection{Pilot-wave equilibrium: A generalisation of quantum equilibrium} \label{pw define}
Consider an ensemble of systems described by a non-normalizable quantum state $\psi(\vec{y})$ with a normalized density of configurations $\rho(\vec{y})$. We want to understand if a physically-meaningful equilibrium density can be defined for the ensemble. In the case of normalizable quantum states, we know that the equilibrium density satisfies the following conditions:\\
1. Entropy maximization: The equilibrium density minimises an appropriately defined $H$-function (the negative of which is maximised).\\
2. Equilibrium stability: The equilibrium density continues to be in equilibrium with time.\\
3. Equivariance: The functional form of the equilibrium density in terms of the quantum state is preserved with time.\\
4. Quantum-mechanical equivalence: The statistical predictions made by the equilibrium density is equal to that predicted by orthodox quantum mechanics for the same quantum state.\\

Let us check whether these conditions can be met in our scenario. Consider the first condition: we typically seek a density $\rho(\vec{y})$ that minimises the $H$-function \cite{valentinI}
\begin{align}
H_q \equiv \int_\mathcal{C} \rho(\vec{y}) \ln \frac{\rho(\vec{y})}{|\psi(\vec{y})|^2} d\vec{y} \label{h}
\end{align}
where the integral is defined over all of configuration space $\mathcal{C} = \{\vec{y}| y_r \in \mathcal{R} \textbf{ } \forall r \}$ and $\mathcal{R}$ is the set of all reals. Equation (\ref{h}) immediately lands us in trouble as it is formally the relative entropy from $\rho(\vec{y})$ to $|\psi(\vec{y})|^2$ -- but $|\psi(\vec{y})|^2$, being non normalizable, is \textit{not} a probability density over $\mathcal{C}$. Therefore, $H_q$ is not a mathematically well-defined relative entropy. \\

Fortunately, it is straightforward to rectify the definition of $H$ for our scenario. We note that, in general, the density $\rho(\vec{y})$ may have support only over a proper subset $\Omega \equiv \{\vec{y}|\rho(\vec{y}) >0\}$ of $\mathcal{C}$. Let us assume that $\Omega$ is a proper subset of $\mathcal{C}$, that is, $\rho(\vec{y})$ has a compact support. We can then treat $|\psi(\vec{y})|^2$ as a probability density over $\Omega$ once appropriately normalized. We define a candidate equilibrium density 
\begin{align}
\rho_{pw}(\vec{y}) \equiv 
\begin{cases}
|\psi(\vec{y})|^2/\mathcal{N} &\text{, for $\vec{y} \in \Omega$}\\
0 &\text{, for $\vec{y} \in \mathcal{C} \setminus \Omega$} \label{sad}
\end{cases}
\end{align}
where $\mathcal{N} \equiv \int_\Omega |\psi(\vec{y})|^2 d\vec{y}$. We then replace $H_q$ by
\begin{align}
H_{pw} \equiv \int_{\mathcal{C}} \rho(\vec{y}) \ln \frac{\rho(\vec{y})}{\rho_{pw}(\vec{y})} d\vec{y} \label{newh}
\end{align}
Note that, since $\rho_{pw}(\vec{y})$ is a valid probability density over $\mathcal{C}$, $H_{pw}$ is a well-defined relative entropy from $\rho(\vec{y})$ to $\rho_{pw}(\vec{y})$. Equation (\ref{newh}) can be written as
\begin{align}
H_{pw} = \int_\mathcal{C} \bigg ( \rho(\vec{y}) \ln \frac{\rho(\vec{y})}{\rho_{pw}(\vec{y})} - \rho(\vec{y}) + \rho_{pw}(\vec{y}) \bigg ) d\vec{y} \label{dadu}
\end{align}
so that the integrand is always non-negative, which implies that the lower bound $H_{pw}^{min} = 0$, which is achieved when $\rho(\vec{y}) = \rho_{pw}(\vec{y})$. Therefore, the newly-defined quantities $\rho_{pw}(\vec{y})$ and $H_{pw}$ together satisfy the first condition set out at the beginning of the subsection.\\

Let us next consider the second condition: does the initial density $\rho(\vec{y}, 0) = \rho_{pw}(\vec{y}, 0)$ evolve to a $\rho(\vec{y}, t)$ that minimises $H_{pw}(t)$? We know that \cite{bohmrelax}, since both $\rho(\vec{y}, t)$ and $|\psi(\vec{y}, t)|^2$ satisfy the same continuity equation, we have 
\begin{align}
\frac{df(\vec{y}, t)}{dt} = \partial_t f(\vec{y}, t) + \vec{\nabla}f(\vec{y}, t) \cdot \vec{v}(\vec{y}, t) = 0 \label{same}
\end{align}
where $f(\vec{y}, t) \equiv \rho(\vec{y}, t)/|\psi(\vec{y}, t)|^2$. Equation (\ref{same}) implies that, given an initial density $\rho_{pw}(\vec{y}, 0)$, we have 
\begin{align}
\rho_{pw}(\vec{y}, t) = |\psi(\vec{y}, t)|^2/\mathcal{N} \text{, if $\vec{y} \in \Omega_t$}\label{what}
\end{align}
where $\Omega_t = \{\vec{y}|\rho_{pw}(\vec{y}, t) >0\}$ is the support of $\rho_{pw}(\vec{y}, t)$. We note that equation (\ref{what}) implies
\begin{align}
\int_{\Omega_t} |\psi(\vec{y}, t)|^2 d\vec{y} = \mathcal{N} \text{ $\forall t$} \label{cons}
\end{align}
The time-dependent $H$-function
\begin{align}
H_{pw}(t) = \int_{\mathcal{C}} \bigg ( \rho(\vec{y}, t) \ln \frac{\rho(\vec{y}, t)}{\rho_{pw}(\vec{y}, t)} - \rho(\vec{y}, t) + \rho_{pw}(\vec{y}, t) \bigg ) d\vec{y} 
\end{align}
remains constant at its lower bound $H_{pw}^{min} (t)= 0$ for the density $\rho(\vec{y}, t) = \rho_{pw}(\vec{y}, t)$. Thus, an initial density that minimises $H_{pw}(0)$ will evolve in time so as to minimise $H_{pw}(t)$ at all times. \\

The third condition, of equivariance, is not directly met as the support $\Omega_t$ is not determined by the quantum state. However, it is clear from (\ref{what}) that the functional form of $\rho_{pw}(\vec{y}, t)$ in terms of $\psi(\vec{y}, t)$ over $\Omega_t$ is invariant with time. We may therefore define the following condition to be pilot-wave invariance: the functional form of the density in terms of the quantum state over its support is invariant with time. Pilot-wave invariance is motivated by the notion of equivariance, and reduces to it in the special case that $\psi$ is normalizable and $\Omega_t = \mathcal{C}$ $\forall t$.\\

Is the fourth condition also met? This condition ceases to make sense in our case, as we are dealing with quantum states that are non-normalizable. Such states are considered unphysical in orthodox quantum mechanics, and the theory provides no experimental probabilities for ensembles with such states. In view of the fact that conditions 1, 2 and 3 (suitably modified) are satisfied, and condition 4 is inapplicable, we may define a density that satisfies only the first three conditions to be in \textit{pilot-wave} equilibrium (as opposed to quantum equilibrium). The terminology makes explicit the fact that $H_{pw}$ quantifies relaxation to an equilibrium density in pilot-wave theory regardless of whether that density reproduces orthodox quantum mechanics, whereas $H_q$ quantifies relaxation to the equilibrium density that reproduces orthodox quantum mechanics. For normalizable states, the notion of pilot-wave equilibrium reduces to quantum equilibrium for the special case when $\Omega = \mathcal{C}$.\\

To conclude, we define a density $\rho(\vec{y}, t)$ with support $\Omega$ to be in pilot-wave equilibrium if and only if
\begin{align}
\rho(\vec{y}, t) = \rho_{pw}(\vec{y}, t) \label{gg}
\end{align}
Clearly, there are infinitely many $\rho(\vec{y}, t)$ that can be in pilot-wave equilibrium, as there are infinitely many subsets $\Omega$ of $\mathcal{C}$. The density $\rho_{pw}(\vec{y}, t)$ minimises the $H$-function
\begin{align}
H_{pw}(t) = \int_\mathcal{C} \rho(\vec{y}) \ln \frac{\rho(\vec{y})}{\rho_{pw}(\vec{y}, t)} d\vec{y}
\end{align}
at all times. If $\rho(\vec{y}, t)$ does not satisfy condition (\ref{gg}), then we define it to be in pilot-wave nonequilibrium. Note that a rescaling $\psi(\vec{y}, t) \to \alpha \psi(\vec{y}, t)$, where $\alpha$ is a complex constant, does not change the equilibrium condition (\ref{gg}), similar to the definition of the velocity field (\ref{vel}). Lastly, we also note that although the concept of pilot-wave equilibrium has been motivated by a consideration of non-normalizable quantum states, it is applicable to normalizable quantum states as well.\\

\subsection{$H$-theorem for relaxation to pilot-wave equilibrium} \label{htheorem}
We now turn to the question whether an arbitrary ensemble density will relax to pilot-wave equilibrium at a coarse-grained level, analogous to relaxation to quantum equilibrium for normalizable states. We show this is indeed the case by proving an $H$-theorem for $H_{pw}$. \\

In the proof for relaxation to classical statistical equilibrium \cite{tallman} or quantum equilibrium \cite{valentinI}, an important role is played by the fact that the exact $H$-function is constant with time. To build an analogous $H$-theorem for pilot-wave equilibrium, our first task then, is to ascertain if $H_{pw}(t)$ is constant with time. From equations (\ref{h}), (\ref{sad}) and (\ref{newh}), the relationship between the two $H$-functions is
\begin{align}
H_{pw}(t) = \ln \mathcal{N}(t) + H_q(t) \label{reln}
\end{align}
Clearly, it is sufficient to prove the constancy of $\mathcal{N}(t)$ to prove that $H_{pw}(t)$ is constant with time. We know, from equation (\ref{cons}), that $\mathcal{N}(t)$ is constant with time if the initial density is in pilot-wave equilibrium. Let us consider an arbitrary initial density $\rho(\vec{y}, 0)$ with support $\Omega_0$ in pilot-wave nonequilibrium, piloted by a non-normalizable state $\psi(\vec{y}, t)$. We also consider the pilot-wave equilibrium density $\rho_{pw}(\vec{y}, 0) = |\psi(\vec{y}, 0)|^2/\mathcal{N}(0)$ over $\Omega_0$, where $\mathcal{N}(0) = \int_{\Omega_0} |\psi(\vec{y}, 0)|^2 d\vec{y}$. As both $\rho(\vec{y}, 0)$ and $\rho_{pw}(\vec{y}, 0)$ are piloted by $\psi(\vec{y}, t)$, they will obey similar continuity equations
\begin{align}
\partial_t \rho(\vec{y}, t) &+ \vec{\nabla} \cdot \big (\rho(\vec{y}, t) \vec{v}(\vec{y}, t) \big ) = 0 \label{oi}\\
\partial_t \rho_{pw}(\vec{y}, t) &+ \vec{\nabla} \cdot \big (\rho_{pw}(\vec{y}, t) \vec{v}(\vec{y}, t) \big ) = 0 \label{kya}
\end{align}
where $\vec{v}(\vec{y}, t)$ is determined by $\psi(\vec{y}, t)$ according to (\ref{vel}). The velocity field $\vec{v}(\vec{y}, t)$ provides the mapping from $\Omega_0 \to \Omega_t$. We also know from equation (\ref{what}) that
\begin{align}
\rho_{pw}(\vec{y}, t) = |\psi(\vec{y}, t)|^2/\mathcal{N}(0) \text{, if $\vec{y} \in \Omega_t$} \label{ttt}
\end{align}
Therefore, the quantity
\begin{align}
\mathcal{N}(t) = \int_{\Omega_t} |\psi(\vec{y}, t)|^2 d\vec{y} = \mathcal{N}(0) \label{cunt}
\end{align}
is in fact constant with time, and we can label it by $\mathcal{N}$. This implies that an arbitrary initial density $\rho(\vec{y}, 0)$ with $\Omega_0$ defined over a region of low (high) $|\psi(\vec{y}, 0)|^2$ will `shrink' (`expand') if it moves to a region of high (low) $|\psi(\vec{y}, t)|^2$. Lastly, equations (\ref{reln}) and (\ref{cunt}) imply that
\begin{align}
\frac{dH_{pw}(t)}{dt} = 0
\end{align}

We are now ready to prove the subquantum $H$-theorem for $H_{pw}$. We first subdivide the configuration space $\mathcal{C}$ into small cells of volume $\delta V$. We then define the coarse-grained quantities
\begin{align}
\overline{\rho(\vec{y}, t)} &\equiv \frac{1}{\delta V} \int_{\delta V} \rho(\vec{y}, t) d\vec{y}\\
\overline{\rho_{pw}(\vec{y}, t)} &\equiv \frac{1}{\delta V} \int_{\delta V} \rho_{pw}(\vec{y}, t) \label{alah}
\end{align}
where the integral $\int_{\delta V} d\vec{y}$ is performed over the cell which contains $\vec{y}$. Clearly, $\overline{\rho(\vec{y}, t)}$ and $\overline{\rho_{pw}(\vec{y}, t)}$ are constant in each cell. We define the quantity
\begin{align}
 g(\vec{y}, t) \equiv 
\begin{cases} 
\rho(\vec{y}, t)/\rho_{pw}(\vec{y}, t)&, \text{if }\vec{y} \in \Omega_t \\
 0&, \text{if }\vec{y} \in \mathcal{C}\setminus \Omega_t
\end{cases} 
\end{align}
and its coarse-grained version $ \overline{g(\vec{y}, t)} \equiv \overline{\rho(\vec{y}, t)}/\overline{\rho_{pw}(\vec{y}, t)}$ if $\vec{y} \in \overline{\Omega_t}$, where $\overline{\Omega_t} \equiv \{\vec{y}|\overline{\rho(\vec{y}, t)} >0\}$ of $\mathcal{C}$. Subtracting (\ref{kya}) from (\ref{oi}) and using the definition of $g(\vec{y}, t)$, we have
\begin{align}
\frac{dg(\vec{y}, t)}{dt} = \partial_t g(\vec{y}, t) + \vec{\nabla}g(\vec{y}, t) \cdot \vec{v}(\vec{y}, t) = 0 \label{same2}
\end{align}
which is analogous to equation (\ref{same}). We define the coarse-grained version of $H_{pw}$ to be
\begin{align}
\overline{H_{pw}(t)} &\equiv \int_{\mathcal{C}} \overline{\rho(\vec{y}, t)} \ln \frac{\overline{\rho(\vec{y}, t)}}{\overline{\rho_{pw}(\vec{y}, t)}} d\vec{y} \label{deaf}\\
&= \int_{\mathcal{C}} \overline{\rho(\vec{y}, t)} \ln \overline{g(\vec{y}, t)} d\vec{y}
\end{align}
Analogous to the $H$-theorems for classical statistical equilibrium \cite{tallman} and for quantum equilibrium \cite{valentinI}, we assume that there is no initial fine-grained structure, that is,
\begin{align}
\rho(\vec{y}, 0) &=  \overline{\rho(\vec{y}, 0)} \label{one} \\
\rho_{pw}(\vec{y}, 0) &= \overline{\rho_{pw}(\vec{y}, 0)} \label{two}
\end{align}

Let us consider
\begin{align}
\overline{H_{pw}(0)} - \overline{H_{pw}(t)} = \int_{\mathcal{C}} \overline{\rho(\vec{y}, 0)} \ln \overline{g(\vec{y}, 0)} d\vec{y} - \int_{\mathcal{C}} \overline{\rho(\vec{y}, t)} \ln \overline{g(\vec{y}, t)} d\vec{y} \label{don}
\end{align}
Using the initial conditions (\ref{one}) and (\ref{two}), and the fact that $H_{pw}(t)$ is constant with time, we can simplify the first term in RHS of (\ref{don}) as
\begin{align}
\int_{\mathcal{C}} \overline{\rho(\vec{y}, 0)} \ln \overline{g(\vec{y}, 0)} d\vec{y}  &= \int_{\mathcal{C}} \rho(\vec{y}, 0) \ln g(\vec{y}, 0) d\vec{y} \\
&= \int_{\mathcal{C}} \rho(\vec{y}, t) \ln g(\vec{y}, t) d\vec{y} \label{hg1}
\end{align}
The second term in RHS of (\ref{don}) can be written as
\begin{align}
\int_{\mathcal{C}} \overline{\rho(\vec{y}, t)} \ln \overline{g(\vec{y}, t)} d\vec{y} = \sum_i \int_{\delta V_i} \overline{\rho(\vec{y}, t)} \ln \overline{g(\vec{y}, t)} d\vec{y}
\end{align}
where the integral over $\mathcal{C}$ has been broken up into integrals over each cell of volume $\delta V$. As $\overline{\rho(\vec{y}, t)}$ and $\overline{\rho_{pw}(\vec{y}, t)}$ are constant over these cells, we can write $\overline{\rho(\vec{y}, t)} = \overline{\rho_i(t)}$, $\overline{\rho_{pw}(\vec{y}, t)} = \overline{\rho_{pwi}(t)}$ and $\overline{g(\vec{y}, t)} = \overline{g_i(t)}$ if $\vec{y}$ belongs to the $i^{th}$ cell. It then follows that
\begin{align}
\sum_i \int_{\delta V_i} \overline{\rho(\vec{y}, t)} \ln \overline{g(\vec{y}, t)} d\vec{y} &= \sum_i \overline{\rho_i(t)} \ln \overline{g_i(t)} \delta V\\
&= \sum_i \overline{\rho_i(t)} \ln \overline{g_i(t)} \frac{\int_{\delta V} \rho(\vec{y}, t) d\vec{y}}{\overline{\rho_i(t)}} \label{jhant}\\
&= \int_{\mathcal{C}} \rho(\vec{y}, t) \ln \overline{g(\vec{y}, t)} d\vec{y} \label{hg2}
\end{align}
where, in equation (\ref{jhant}), we have used the relation (\ref{one}). Using (\ref{hg1}) and (\ref{hg2}), we can rewrite (\ref{don}) as
\begin{align}
\overline{H_{pw}(0)} - \overline{H_{pw}(t)} &= \int_{\mathcal{C}} \rho(\vec{y}, t) \ln \frac{g(\vec{y}, t)}{\overline{g(\vec{y}, t)}} d\vec{y}\\
&= \int_{\mathcal{C}} \rho_{pw}(\vec{y}, t) g(\vec{y}, t)\ln \frac{g(\vec{y}, t)}{\overline{g(\vec{y}, t)}} d\vec{y} \label{rew}
\end{align}

We note that
\begin{align}
\int_{\mathcal{C}} \rho_{pw}(\vec{y}, t)\overline{g(\vec{y}, t)}d\vec{y} &= \sum_i \int_{\delta V_i} \rho_{pw}(\vec{y}, t)\frac{\overline{\rho(\vec{y}, t)}}{\overline{\rho_{pw}(\vec{y}, t)}} d\vec{y}\\
&= \sum_i \frac{\overline{\rho_i(t)}}{\overline{\rho_{pwi}(t)}} \int_{\delta V_i} \rho_{pw}(\vec{y}, t) d\vec{y}\\
&= \sum_i \overline{\rho_i(t)} \delta V\\
&= \int_{\mathcal{C}} \overline{\rho(\vec{y}, t)}d\vec{y} = 1 \label{1l}
\end{align}

Using (\ref{1l}), we can rewrite equation (\ref{rew}) as 
\begin{align}
\overline{H_{pw}(0)} - \overline{H_{pw}(t)} = \int_{\mathcal{C}} \rho_{pw}(\vec{y}, t) \bigg ( g(\vec{y}, t)\ln \frac{g(\vec{y}, t)}{\overline{g(\vec{y}, t)}} - g(\vec{y}, t) + \overline{g(\vec{y}, t)} \bigg ) d\vec{y} \label{fck}
\end{align}
Using the identity $x\ln(x/y) -x + y \geq 0$ for all real $x, y$, it is then clear from equation (\ref{fck}) that $\overline{H_{pw}(0)} - \overline{H_{pw}(t)} \geq 0$. We have, therefore, proven an $H$-theorem for $\overline{H_{pw}(t)}$, subject to assumptions similar to those assumed for relaxation to quantum equilibrium. 

\subsection{Relationship between relaxation to pilot-wave equilibrium and to quantum equilibrium} \label{num}
Although the $H$-theorem for $H_{pw}$ gives the theoretical basis for relaxation to pilot-wave equilibrium, we need numerical evidence to determine whether relaxation in fact occurs. There exists a large body of results in the literature on the numerical evidence for relaxation to quantum equilibrium for normalizable states. It is, therefore, of interest to understand the relation between relaxation to pilot-wave equilibrium for non-normalizable states and relaxation to quantum equilibrium for normalizable states, if any.\\

We begin by noting that equation (\ref{alah}) can be written as 
\begin{align}
\overline{\rho_{pw}(\vec{y}, t)} &= \frac{1}{\delta V} \int_{\delta V} \frac{|\psi(\vec{y}, t)|^2}{\mathcal{N}}\\
&= \frac{\overline{|\psi(\vec{y}, t)|^2}}{\mathcal{N}} \label{refo}
\end{align}
where $\overline{|\psi(\vec{y}, t)|^2} \equiv \int_{\delta V} |\psi(\vec{y}, t)|^2/\delta V$ and $\vec{y} \in \Omega_t$. From equations (\ref{deaf}) and (\ref{refo}), we can then derive
\begin{align}
\overline{H_{pw}(t)} = \overline{H_{q}(t)} + \ln \mathcal{N} \label{reln2}
\end{align}
where 
\begin{align}
\overline{H_{q}(t)} \equiv \int_{\mathcal{C}} \overline{\rho(\vec{y}, t)} \ln \frac{\overline{\rho(\vec{y}, t)}}{\overline{|\psi(\vec{y}, t)|^2}} d\vec{y}
\end{align}
It is clear from (\ref{reln2}) that the lower bound of $\overline{H_{q}(t)}$ is $H_{q}^{min} = -\ln \mathcal{N}$, corresponding to pilot-wave equilibrium $H_{pw}^{min} = 0$. The relationship (\ref{reln2}) implies that a study of the behaviour of $\overline{H_{q}(t)}$ is equivalent to that of $\overline{H_{pw}(t)}$. It now remains to recast this study in terms of normalizable states.\\

Consider the non-normalizable quantum state $\psi(\vec{y}, t) = \sum_{j=1}^n c_j(t) \prod_{g=1}^N \psi^{K_j^g}(y_g)$ from equation (\ref{khur}). We know that the velocity field $v_r(\vec{y}) \sim 1/y_r^2$ at large $\pm y_r$. Suppose a number $L$ sufficiently large such that $v_r(\vec{y})$ is very small at $y_r = \pm L$, then an initial distribution $\rho(\vec{y}, 0)$ localised in the region $|y_r| \leq L$ cannot escape to $|y_r| > L$ for an arbitrarily long time (depending on the value of $L$ chosen). This implies that we effectively need only $v_r(\vec{y})$ for $y_r \in (-L, +L)$ to know how $\rho(\vec{y}, t)$ evolves in the $y_r$ direction. We can utilise this feature of the velocity field to define a normalizable quantum state with the same velocity field in the region $y_r \in (-L, +L)$ as that of the non-normalizable quantum state.\\

Let us define the normalizable quantum state 
\begin{align}
\psi_n(\vec{y}, t) \equiv \prod_{r=1}^N e^{-\theta(y_r - L) (y_r - L)^{2m}} e^{-\theta(-y_r - L) (y_r + L)^{2m}}  \psi(\vec{y}, t)\label{n}
\end{align}
where $\theta(x)$ is the Heaviside-step function, $m$ is a positive integer and $L$ is a very large constant such that $v_r(\vec{y})$ is very small at $y_r = \pm L$ for all $r \in \{1, 2, ...N\}$. We know that $\psi_n(\vec{y}, t)$ is normalizable as $\psi^{K_j^g}(\vec{y}) \sim e^{y_r^2/2}$ at large $\pm y_r$ for all $r \in \{1, 2, ...N\}$. Clearly, we can replace $\psi(\vec{y}, t)$ by $\psi_n(\vec{y}, t)$ to evolve $\rho(\vec{y}, t)$ if $\rho(\vec{y}, 0)$ has an initial support $\Omega_0 \subset \Lambda \equiv \{\vec{y}| y_r \in (-L, +L) \textbf{ }\forall \textbf{ } r\}$. The evolution of $\psi_n(\vec{y}, t)$ itself is non-unitary as $e^{-i\hat{H}t/\hbar} \psi_n(\vec{y}, 0) \neq e^{-i\hat{H}t/\hbar} \psi(\vec{y}, 0)$. This is because $\psi_n(\vec{y}, 0)$ is numerically, but not \textit{functionally}, equal to $\psi(\vec{y}, 0)$ in the subset $\Lambda$. Therefore, we can study relaxation to pilot-wave equilibrium using normalizable states, but doing so would require non-unitary dynamics. A complete relaxation to pilot-wave equilibrium $H_{pw}^{min} = 0$ would correspond to a partial relaxation to quantum equilibrium $H_{q}^{min} = -\ln \mathcal{N}$ (see Fig. 4). \\

\begin{figure}
\centering
\includegraphics[scale=0.45]{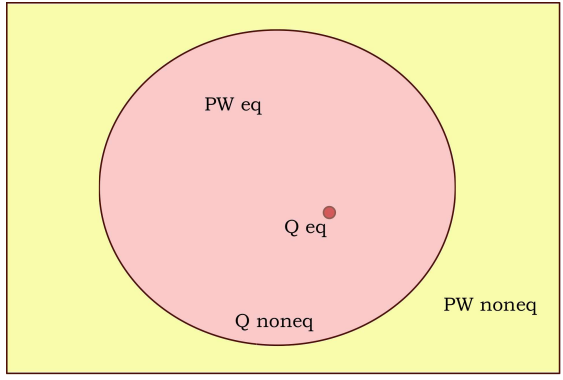}
\caption{Schematic illustration of the relationship between quantum equilibrium (Q eq) and the notion of pilot-wave equilibrium (PW eq) introduced in this paper. Given a normalizable quantum state $\psi$, there is only a single density $\rho_q = |\psi|^2$ that is defined to be in quantum equilibrium (depicted as the dark red dot). On the other hand, there is an infinite number of densities $\rho_{pw}$ that are in pilot-wave equilibrium (depicted as the light red region), corresponding to different subsets $\Omega$ of the configuration space. Quantum equilibrium is a special case of pilot-wave equilibrium as depicted. For non-normalizable states, there is no density in quantum equilibrium (there is no red dot) but there are densities in pilot-wave equilibrium.}
\end{figure}

\section{Theoretical and experimental implications} \label{impli}
In this section, we sketch the theoretical and experimental implications of our work. Although we have focused on the harmonic oscillator, the general approach adopted in this paper and the notion of pilot-wave equilibrium introduced are not exclusive to the harmonic oscillator. Therefore, where applicable, we discuss the implications in the broader context of non-normalizable quantum states with a normalized density of configurations.\\

\subsection{Non-relativistic quantum theory}

\subsubsection{Experimental observation of continuous-energy eigenstates} \label{nonrel}
We have seen that pilot-wave theory gives a physical interpretation for non-normalizable harmonic oscillator states as bound states. However, such states have continuous energies and have never been experimentally observed. Does this directly falsify pilot-wave theory in favour of orthodox quantum mechanics?\\

We first note that unitarity imposes restrictions on preparation of non-normalizable states in a laboratory. This is because, if the initial joint quantum state of the preparation apparatus (including all the atoms of all the equipments etc.) is normalizable, then the joint quantum state will remain normalizable after the preparation is completed. The argument can be repeated to conclude that non-normalizable states can be potentially detected today only if there existed non-normalizable states in the early universe.\\

Consider an atom in the early universe in a non-normalizable eigenstate $\psi^K(\vec{y})$, where $K$ is continuous. The atom will, in general, be subject to small perturbations $\delta V(\Vec{y}, t)$ across the universe. It can be shown, from time-dependent perturbation theory, that the quantum state will evolve as 
\begin{align}
    \psi(\Vec{y}, t) = e^{-i\hat{H}_0 t/\hbar} \psi^K(\vec{y}) - \frac{ie^{-i\hat{H}_0 t/\hbar}}{\hbar}\int_0^t dt' e^{i\hat{H}_0 t'/\hbar} \delta V(\Vec{y}, t') e^{-i\hat{H}_0 t'/\hbar} \psi^K(\Vec{y}) + \mathcal{O}(\delta V^2) \label{pert}
\end{align}
up to first order in $\delta V$, where $\hat{H}_0$ is the unperturbed Hamiltonian of the atom. Note that, as the Dyson series does not assume state normalizability \cite{sakurai}, equation (\ref{pert}) is valid for $\psi^K(\Vec{y})$. Let us consider realistic perturbations $\delta V(\Vec{y}, t')$ that are small and localised in space. That is, suppose the perturbations are of the approximate form
\begin{align}
    \delta V(\Vec{y}, t') = \sum_{n=1}^N e^{\frac{-|\vec{y}-\vec{y}_n(t')|^4}{\sigma_n}}
\end{align}
so that they rapidly fall off around $\vec{y}_n(t')$. Then, using the fact that $\psi^K(\vec{y})$ is an eigenstate, we can write the integrand in (\ref{pert}) as 
\begin{align}
    e^{i\hat{H}_0 t'/\hbar} \bigg( e^{-iE_K t'/\hbar} \delta V(\Vec{y}, t') \psi^K(\Vec{y}) \bigg) = \sum_j e^{i(E_j -E_K) t'/\hbar} c_j(t') \psi^j(\vec{y})
\end{align}
as $\delta V(\Vec{y}, t')\psi^K(\Vec{y})$ is square integrable (although $\psi^K(\Vec{y})$ is not) and can be expanded in terms of the normalizable eigenstates $\psi^j(\vec{y})$ of $H_0$. Note that a perturbation $\delta V(\Vec{y}, t')$ arbitrarily distant from the atom is sufficient to make $\delta V(\Vec{y}, t')\psi^K(\Vec{y})$ square integrable, given that $\delta V(\Vec{y}, t')$ falls off rapidly. Therefore, for realistic perturbations equation (\ref{pert}) becomes
\begin{align}
    \psi(\Vec{y}, t) = e^{-iE_K t/\hbar} \psi^K(\vec{y}) - \frac{ie^{-i\hat{H}_0 t/\hbar}}{\hbar}\int_0^t dt' \sum_j e^{i(E_j -E_K) t'/\hbar} c_j(t') \psi^j(\vec{y}) + \mathcal{O}(\delta V^2) \label{pertu}
\end{align}
so that the quantum state becomes a superposition of the non-normalizable $\psi^K(\vec{y})$ and the normalizable $\psi^j(\vec{y})$'s. If the atom now interacts strongly with the environment to cause an effective energy-measurement, then the possible eigenvalues are the discrete energies $E_j$ as well as the continuous energy $E_K$. Using the von-Neumann measurement\cite{vonN} Hamiltonian $\hat{H}_I = g \hat{E}_{\vec{y}}\otimes \hat{p}_x$, we can represent the combined state of the atom and an idealised pointer variable after such a measurement to be
\begin{align}
    \Psi(\Vec{y}, x, t) = \sum_n a_n \phi(x - \frac{gtE_n}{\hbar^2},0)\psi^n(\vec{y})
\end{align}
where $g$ is the interaction constant, $\phi(x - \frac{gtE_n}{\hbar^2},0)$ is the pointer state, and $\psi^n(\vec{y})$ is used to represent both $\psi^j(\vec{y})$ and $\psi^K(\vec{y})$ in the superposition (\ref{pertu}). The probabilities will not be given by the Born rule as $\Psi(\Vec{y}, x, t)$ is non-normalizable, but will have to be computed from the normalized probability density $\rho(\vec{y}, t)$. Note that decoherence will effectively occur as long as the pointer wavefunction $\phi(x - \frac{gtE_n}{\hbar^2},0)$ is normalizable. Further interactions with macroscopic bodies will cause further decoherence \cite{bohm2}, so that the measurement will be effectively irreversible as for normalizable quantum states. \\

Therefore the atom, on account of perturbations and interactions with environment, may transition to a normalizable energy eigenstate. In that case, the total quantum state $\Psi(\Vec{y}, x, t)$ will remain non normalizable but the system configuration will enter an effectively-decohered normalizable branch. After $N$ such measurements, the fraction that remains in the non-normalizable branch will be given by
\begin{align}
    f(N) = \Pi_{j=1}^N (1-\epsilon_j)
\end{align}
where the fraction lost to the normalizable branches in the $j^{th}$ measurement is labelled by $\epsilon_j$. Clearly, $f(N) \to 0$ as $N \to \infty$ unless $\epsilon_j = 0$ $\forall j > N_0$ where $N_0$ is some positive integer. The condition $\epsilon_j = 0$ $\forall j > N_0$ is possible if the initial density, the initial joint quantum state of the atom and the idealised measurement apparatus, and the perturbations are so finely tuned that the configuration density remains completely in the non-normalizable branch for all $j > N_0$. Without such fine tuning, the probability of the atom remaining in $\psi^K(\Vec{y})$ becomes tiny after a sufficiently long time corresponding to a large $N$. Note the key role played by perturbations here as they continuously add superpositions of normalizable eigenstates to the total quantum state. Therefore, we would not in general expect non-normalizable states in the early universe to have survived to the present time. Further technical work is required to ascertain the survival timescales for various non-normalizable states and perturbations.

\subsubsection{Signalling and pilot-wave equilibrium}
We know that no-signalling is generally violated in quantum nonequilibrium \cite{valentinII}. Given that quantum equilibrium (when applicable) is a special case of pilot-wave equilibrium, it is of interest to understand the signalling behaviour of ensembles in pilot-wave equilibrium. This is important to understand whether non-normalizable states in pilot-wave equilibrium are no-signalling. Below, we show that no-signalling is violated generally in pilot-wave equilibrium. \\

Consider an initial two-particle entangled quantum state $\psi(y_1, y_2, 0)$, where the two particles are located in space-like separated wings. Suppose an initial density with the support $\Omega_0 \equiv \{(y_1, y_2)|y_1\in (Y_1, Y_1 +\delta y_1) , y_2\in (Y_2, Y_2 +\delta y_2) \}$ where $\delta y_1, \delta y_2$ are very small. Then $\mathcal{N} = \int_{\Omega_0} |\psi(y_1, y_2, 0)|^2 dy_1dy_2  \approx |\psi(Y_1, Y_2, 0)|^2 \delta y_1\delta y_2$. The density $\rho(y_1, y_2, 0) \equiv |\psi(y_1, y_2, 0)|^2/\mathcal{N} \approx 1/ \delta y_1\delta y_2$ on $\Omega_0$ is in pilot-wave equilibrium by definition. \\

Suppose $\psi(y_1, y_2, t)$ evolves under the Hamiltonian $\hat{H} = \hat{H}_1\otimes \hat{I} + \hat{I} \otimes \hat{H}_2$. The question is whether the marginal density of $y_1$ is affected by the distant local Hamiltonian $\hat{H}_2$ under the control of the experimenter at the second wing. We know that, since $\psi(y_1, y_2, t)$ is entangled, the velocity of the first particle $v_1(y_1, y_2, t)$ will depend on $y_2$ and thereby on $\hat{H}_2$. Furthermore, in the limit $\delta y_1, \delta y_2 \to 0$, $\rho(y_1, y_2, 0) = \delta(y_1 - Y_1) \delta(y_2 - Y_2)$ and the initial marginal density of the first particle will be $\rho(y_1, 0) = \delta(y_1 - Y_1)$. It is then clear that, since $v_1(y_1, y_2, t)$ depends on $\hat{H}_2$ and $\rho(y_1, 0)$ contains only the point $Y_1$, $\rho(y_1, t) = \delta(y_1 - Y_1(t))$ will depend on $\hat{H}_2$. The statistics of a position measurement performed at the first wing at time $t$ will then depend on the Hamiltonian chosen by the experimenter at the second wing. We conclude that, in general, correlations generated by an ensemble in pilot-wave equilibrium are signalling, unless the ensemble is also in quantum equilibrium. As there is no notion of quantum equilibrium for non-normalizable states, we conclude that non-normalizable states generate signalling correlations in general. 

\subsection{Quantum field theory}
We know that quantum fields can often be treated as a collection of harmonic oscillators \cite{macha}. For illustration, let us consider the pilot-wave treatment \cite{debbqft} of a free, massless real scalar field $\phi(\vec{x}, t)$ on a flat expanding space-time, with the Lagrangian density $\mathcal{L} = \big (a^3 \dot{\phi}^2 - a (\nabla \phi)^2\big )/2$, where $a=a(t)$ is the scale factor and $c =1$ for simplicity. The functional Schrodinger equation for this system is
\begin{align}
\sum_{\vec{k}, r}\big (\frac{1}{2a^3}\pi_{\vec{k}, r}^2 + \frac{ak^2}{2}q_{\vec{k},r}^2 \big )\psi = i\frac{\partial \psi}{\partial t} \label{func}
\end{align}
where $\psi = \psi(\{q_{\vec{k}, r}\}, t)$ is the quantum state defined over the configuration space $\{q_{\vec{k}, r}\} \equiv (q_{\vec{k}_1, r}, q_{\vec{k}_2, r}, q_{\vec{k}_3, r}...)$, $q_{\vec{k}, r}$ $(r = 1, 2)$ are real variables related to the Fourier-transform of $\phi(\vec{x}, t)$ by 
\begin{align}
\phi(\vec{k}, t) \equiv \frac{1}{(2\pi)^{3/2}} \int \phi(\vec{x}, t) e^{-i\vec{k}\cdot\vec{x}} d\vec{x} = \frac{\sqrt{V}}{(2\pi)^{3/2}} \big ( q_{\vec{k}, 1}(t) + i q_{\vec{k}, 2}(t)\big )
\end{align}
and $\pi_{\vec{k}, r} \equiv \partial (\int \mathcal{L} d\vec{x})/\partial \dot{q}_{\vec{k}, r} = a^3 \dot{q}_{\vec{k}, r}$ is the canonical momentum. Here $V$ is the box-normalization volume. Note that equation (\ref{func}) assumes a regularization so that a finite (but arbitrarily large) number of $\vec{k}$ can be considered. \\

Equation (\ref{func}) clearly shows that $\phi(\vec{x}, t)$ can be treated as a collection of independent harmonic oscillators in the Fourier space. Notably, although the field $\phi(\vec{x}, t)$ is assumed to have a Fourier-transform, we need not make the same assumption about $\psi(\{q_{\vec{k}, r}\}, t)$ which is piloting $\phi(\vec{x}, t)$. Therefore, we can consider the non-normalizable solutions to (\ref{func}) explored in this paper. Such solutions may have implications in cosmological settings \cite{debbqft, teenv}.

\subsection{Quantum gravity}
It is well known that non-normalizable quantum states are often encountered in quantum gravity \cite{witten003, allin, randi}. Such states are also encountered when pilot-wave dynamics is formulated on shape space, where a different approach to the problem of non-normalizability from a pilot-wave perspective has been explored \cite{durrsd}. Recently, Valentini has argued for a pilot-wave approach to quantum gravity where statistical predictions are derived from a normalized configuration density \cite{chori}. This is close to the approach adopted in our work, but there are several important differences. It is useful to discuss the implications of our work for quantum gravity in the context of ref. \cite{chori}.\\

First, ref. \cite{chori} argues that there is no physical equilibrium density for non-normalizable quantum states, on the basis that the lower bound of $H_q$ diverges to $-\infty$. However, this argument has multiple flaws. Firstly, the lower bound of $H_q$ diverges only in the particular case where the support $\Omega$ of the configuration density is the entire configuration space $\mathcal{C}$, that is $\Omega = \mathcal{C}$. For all other cases the lower bound of $H_q$ is $H_q^{min} = -\ln \mathcal{N}$, as can be seen from equation (\ref{reln}). Secondly, we have argued that, for non-normalizable quantum states, the notion of quantum equilibrium must be replaced by the more general notion of pilot-wave equilibrium. Correspondingly, $H_q$ must be replaced by $H_{pw}$ to define a physical equilibrium density. Therefore, our results imply that some form of the Born rule arises as a physical equilibrium density for non-normalizable states. \\

Second, ref. \cite{chori} has emphasised that non-normalizability of the quantum state is due to the ``deep physical reason'' that the Wheeler-DeWitt equation on configuration space has a Klein-Gordon-like structure. In our approach on the other hand, there is no special role played by the structure of any particular equation. We have argued that non-normalizability is intrinsic to pilot-wave theory -- only a normalized configuration density is needed to obtain statistical predictions. The quantum state, which defines the evolution of the configuration, need not be normalizable. Therefore, non-normalizable quantum states naturally follow from the first principles of the theory and the structure of the Wheeler-Dewitt equation can only play a technical role. This implies that non-normalizable solutions to the Schrodinger equation or Dirac equation are as valid from a pilot-wave perspective, where applicable, as that to the Wheeler-DeWitt equation. \\

\section{Discussion} \label{disc}
We have discussed some of the implications of our work in the previous section. However, the list of implications is necessarily inexhaustive as the normalizability constraint is ubiquitous in orthodox quantum mechanics. It would, for example, be interesting to study non-normalization solutions to the Schrodinger equation for other systems, say the Hydrogen atom, or to the Dirac equation. An important result of our work is that the non-normalizable harmonic-oscillator solutions are bound states, in the sense that the pilot-wave velocity field $v_y \to 0$ at large $\pm y$. It is important to figure out the general conditions under which the pilot-wave velocity field has this behaviour. Another important result is that perturbations and interactions make non-normalizable states unstable, in the sense that the system configuration becomes overwhelmingly likely with time to be in a normalizable branch of the total quantum state. Lastly, it remains unclear how to construct a well-defined basis for such states. \\

We note that, according to our work, the explanation for quantization given by pilot-wave theory is drastically different from that of quantum mechanics. Quantization in quantum mechanics arises from the axiom of Born rule, whereas in pilot-wave theory quantization is an emergent phenomenon that arises from the instability of non-normalizable states due to perturbations and environmental interactions. In this sense, the status of non-normalizable states in the theory may be said to be analogous to that of non-equilibrium ensembles as \textit{a)} the conceptual structure of the theory allows the logical possibility of both non-normalizable states and non-equilibrium densities, and \textit{b)} the theory also possesses the internal logic necessary to explain why we do not observe either of them in present-day laboratories. \\

We note that the $H$-theorem does not by itself prove that relaxation to pilot-wave equilibrium occurs, but provides a general mechanism to understand how equilibrium is approached, similar to the status of the generalized $H$-theorem in classical statistical mechanics \cite{tallman}. Whether relaxation in fact occurs in finite time, if it is monotonic etcetera significantly depend on whether the velocity field yields sufficient mixing. It is well-known in the literature on relaxation in pilot-wave theory \cite{royalvale, nick, teenv} that the velocity field varies rapidly around nodes (if they exist) and thereby causes efficient relaxation in general. Therefore, future numerical simulations using superpositions of non-normalizable eigenstates can provide evidence whether relaxation to pilot-wave equilibrium indeed occurs, similar to relaxation to quantum equilibrium for normalizable states. It is useful to note here that the boundedness of the solutions ensures that the support $\Omega_t$ does not necessarily become filamentous with time. For example, if $\Omega_0$ is sufficiently large to cover the region around the origin and $|\psi(\vec{y},t)|^2$ is very large near its boundary $\partial \Omega_0$, then $\partial \Omega_t$ will remain effectively static as the radial velocity field will be very small in that region. Lastly, we note that the coarse-graining cells do not become filamentous as they do not evolve with time, unlike the configuration density.\\

From a historical perspective, we know that the initial conditions of pilot-wave theory have usually been so restricted as to reproduce orthodox quantum mechanics. An important departure was made when nonequilibrium densities were taken seriously in the theory, and the notion of quantum equilibrium was defined \cite{valentinI, valentinII}. But the notion of quantum equilibrium is still restrictive as it assumes that a density in equilibrium always reproduces orthodox quantum mechanics. The notion of pilot-wave equilibrium makes one further step, in which this restriction is jettisoned. Therefore, generalising the notion of quantum equilibrium to pilot-wave equilibrium may be seen as a logical step towards treating pilot-wave theory as a theory in its own right, instead of as a hidden-variable reformulation of orthodox quantum mechanics. \\

It may appear that the restriction of the configuration density to compact supports limits the physical applicability of pilot-wave equilibrium. However, this is incorrect as we can always approximate a density with global support up to arbitrary accuracy using a density with compact support. This can be done by defining an arbitrarily small but finite cut-off parameter $\epsilon << 1$ so that if the global density $\rho_g (\Vec{y}) \leq \epsilon$ at a particular point $\Vec{y}$ on the configuration space, we define the compact density $\rho_c (\Vec{y}) \equiv 0$, where $\rho_c (\Vec{y}) \equiv \rho_g (\Vec{y})$ (up to normalization) at all other $\Vec{y}$. Further, global supports imply arbitrarily small probabilities that cannot be empirically verified and are, therefore, mathematical idealisations. For example, a Hydrogen atom in a lab on Earth has a finite but arbitrarily small probability of being found, in a position measurement, arbitrarily far away from the Earth. But observing such an extremely tiny probability trillions of light years away would take many times more than the current age of the universe in any realistic experimental setup.\\

There are several implications of our work for pilot-wave theory. First, our work suggests a constraint on the pilot-wave velocity field. We know that the pilot-wave velocity is not uniquely defined as one can always add a divergence-free term to the current. In the context of non-normalizable states, the velocity field plays the important role of determining whether a given state is bounded. Therefore, it seems reasonable to impose the constraint that the addition of divergence-free term to the current does not affect the boundedness of the state. That is, if the (usually defined) pilot-wave velocity field $v_y = j_y/|\psi|^2$ goes to $0$ at $y \to \pm \infty$, then this behaviour must be preserved on modifying $\vec{j} \to \vec{j} + \vec{\nabla} \times \vec{A}$. It would be interesting to figure out the class of possible $\vec{A}$ that satisfy this property. Second, our work may help in distinguishing pilot-wave theory from orthodox quantum mechanics and other realist interpretations of quantum mechanics. For example, some authors have claimed that the system configuration in pilot-wave theory is superflous and the theory is actually a many-worlds theory in disguise \cite{d96eusch, zeh99, brownwal}. As we have seen, however, the existence of a configuration density in the theory makes it possible to extract statistical predictions from non-normalizable quantum states. Therefore, the interpretation of non-normalizable states may turn out to be a crucial difference between the two theories. Third, we note that the notion of pilot-wave equilibrium, although introduced in the context of non-normalizable quantum states, is equally applicable to normalizable quantum states. It would be of interest to figure out whether densities partially relaxed to quantum equilibrium in previous numerical simulations have in fact relaxed to pilot-wave equilibrium. Lastly, our results imply that a unitary evolution involving non-normalizable states is dynamically equivalent to a corresponding non-unitary evolution involving appropriate normalizable states. This suggests that non-unitary evolution in some applications of orthodox quantum mechanics may in fact be an artefact of insistence on state normalizability. This also implies that, for normalizable states, unitary evolution is not necessary for relaxation to pilot-wave equilibrium.\\

Our work also has implications for the $\psi$-ontic versus $\psi$-epistemic debate \cite{harrikens, pbr, leifer}. Non-normalizable quantum states do not make sense from a $\psi$-epistemic viewpoint, in which the role of the quantum state is to define probabilities. If the existence of non-normalizable quantum states is proved experimentally, or if such states are found to be crucial in fields like quantum cosmology or quantum gravity, then it would be difficult to argue in favour of $\psi$-epistemicity. We note that, once pilot-wave equilibrium is reached at a coarse-grained level, then the relation $\overline{\rho(\vec{y}, t)} = \overline{|\psi(\vec{y}, t)|^2}/\mathcal{N}$ on $\Omega_t$ suggests how a $\psi$-epistemic interpretation may emerge at an effective level from an underlying $\psi$-ontic theory.\\

We conclude that pilot-wave theory naturally suggests consideration of the possibility of non-normalizable quantum states, which we have studied for the case of harmonic oscillator. Such states have a physically-meaningful notion of an equilibrium density. We have argued that quantization emerges in pilot-wave theory due to the instability of non-normalizable states to perturbations and environmental interactions. Further work is needed to determine whether such states actually exist in nature.\\~\\
\textbf{Acknowledgements} I am thankful to Matt Leifer for encouragement and several helpful discussions. I am also thankful to Siddhant Das and Tathagata Karmakar for helpful discussions. The author was supported by a fellowship from the Grand Challenges Initiative at Chapman University. 

\bibliographystyle{bhak}
\bibliography{bib}

\begin{thebibliography}{10}

\bibitem{bell}
J.~S. Bell.
\newblock {\em {Speakable and unspeakable in quantum mechanics: Collected
  papers on quantum philosophy}}.
\newblock {Cambridge Univ. Press}, {2004}.

\bibitem{solventini}
G.~Bacciagaluppi and A.~Valentini.
\newblock {\em {Quantum theory at the crossroads: reconsidering the 1927 Solvay
  conference}}.
\newblock {Cambridge Univ. Press}, 2009.

\bibitem{bohm1}
D.~Bohm.
\newblock {A suggested interpretation of the quantum theory in terms of
  ``hidden" variables. I}.
\newblock {\em {Phys. Rev.}}, 85(2), 1952.

\bibitem{bohm2}
D.~Bohm.
\newblock {A suggested interpretation of the quantum theory in terms of
  ``hidden" variables. II}.
\newblock {\em {Phys. Rev.}}, {85}({2}), {1952}.

\bibitem{bohm53}
D.~Bohm.
\newblock {Comments on an article of Takabayasi conserning the formulation of
  quantum mechanics with classical pictures}.
\newblock {\em Prog. Theor. Phys}, 9(3):273--287, 1953.

\bibitem{bohmbook2}
D.~Bohm and B.~J. Hiley.
\newblock {\em The undivided universe: An ontological interpretation of quantum
  theory}.
\newblock Routledge, London, UK, 1993.

\bibitem{valentiniphd}
A.~Valentini.
\newblock {\em {On the pilot-wave theory of classical, quantum and subquantum
  physics}}.
\newblock SISSA, 1992.

\bibitem{struyvefields}
W.~Struyve.
\newblock {Pilot-wave theory and quantum fields}.
\newblock {\em Rep. Prog. Phys}, 73(10):106001, 2010.

\bibitem{durr14}
D.~D{\"u}rr, S.~Goldstein, T.~Norsen, W.~Struyve, and N.~Zangh{\`\i}.
\newblock {Can Bohmian mechanics be made relativistic?}
\newblock {\em Proc. R. Soc. A}, 470, 2014.

\bibitem{valentiniastro}
A.~Valentini.
\newblock {Astrophysical and cosmological tests of quantum theory}.
\newblock {\em {J. Phys. A}}, {40}({12}), {2007}.

\bibitem{pintu19}
N.~Pinto-Neto and W.~Struyve.
\newblock {Bohmian quantum gravity and cosmology}.
\newblock In {\em Applied Bohmian Mechanics}, pages 607--664. Jenny Stanford
  Publishing, 2019.

\bibitem{pk20}
A.~Kandhadai and A.~Valentini.
\newblock {Mechanism for nonlocal information flow from black holes}.
\newblock {\em Int. J. Mod. Phys A}, 35(06):2050031, 2020.

\bibitem{teenv}
A.~Valentini.
\newblock {Foundations of statistical mechanics and the status of the Born rule
  in de Broglie-Bohm pilot-wave theory}.
\newblock In {\em Statistical Mechanics and Scientific Explanation:
  Determinism, Indeterminism and Laws of Nature}. World Scientific, 2020.

\bibitem{bohmrelax}
D.~Bohm.
\newblock {Proof that probability density approaches| $\psi$| 2 in causal
  interpretation of the quantum theory}.
\newblock {\em Phys. Rev}, 89(2):458, 1953.

\bibitem{killer}
J.~B. Keller.
\newblock {Bohm's interpretation of the quantum theory in terms of" hidden"
  variables}.
\newblock {\em Phys. Rev}, 89(5):1040, 1953.

\bibitem{pauli53}
P.~Wolfgang.
\newblock Remarques sur le probl{\`e}me des param{\`e}tres cach{\'e}s dans la
  m{\'e}canique quantique et sur la th{\'e}orie de l'onde pilote.
\newblock In {\em Louis de Broglie: physicien et penseur}. Albin Michel Paris,
  1953.

\bibitem{bohm54}
D.~Bohm and J.-P. Vigier.
\newblock {Model of the causal interpretation of quantum theory in terms of a
  fluid with irregular fluctuations}.
\newblock {\em {Phys. Rev.}}, 1954.

\bibitem{valentinI}
A.~Valentini.
\newblock {Signal-locality, uncertainty, and the subquantum H-theorem. I}.
\newblock {\em {Phys. Lett. A}}, {156}({1-2}), {1991}.

\bibitem{royalvale}
A.~Valentini and H.~Westman.
\newblock {Dynamical origin of quantum probabilities}.
\newblock {\em Proc. R. Soc. A}, 461(2053):253--272, 2005.

\bibitem{kodamaginal}
H.~Kodama.
\newblock {Holomorphic wave function of the universe}.
\newblock {\em Phys. Rev. D}, 42(8):2548, 1990.

\bibitem{witten003}
E.~Witten.
\newblock {A note on the Chern-Simons and Kodama wavefunctions}.
\newblock {\em arXiv gr-qc/0306083}, 2003.

\bibitem{allin}
L.~Smolin.
\newblock {Quantum gravity with a positive cosmological constant}.
\newblock {\em arXiv hep-th/0209079}, 2002.

\bibitem{frizzy}
L.~Bers, F.~John, and M.~Schechter.
\newblock {\em {Partial differential equations}}.
\newblock American Mathematical Soc., 1964.

\bibitem{hassanahi}
S.~Hassani.
\newblock {\em {Mathematical methods: for students of physics and related
  fields}}, volume~2.
\newblock Springer, 2009.

\bibitem{tallman}
R.~C. Tolman.
\newblock {\em {The principles of statistical mechanics}}.
\newblock Oxford Univ. Press, 1938.

\bibitem{sakurai}
J.~J. Sakurai and S.~F. Tuan.
\newblock {Modern quantum mechanics, revised edition}, 1994.

\bibitem{vonN}
J.~Von~Neumann.
\newblock {\em {Mathematical foundations of quantum mechanics}}.
\newblock Princeton Univ. Press, 1955.

\bibitem{valentinII}
A.~Valentini.
\newblock {Signal-locality, uncertainty, and the subquantum H-theorem. II}.
\newblock {\em {Phys. Lett. A}}, {158}({1-2}), {1991}.

\bibitem{macha}
F.~Mandl and G.~Shaw.
\newblock {\em {Quantum field theory}}.
\newblock John Wiley \& Sons, New York, USA, 2010.

\bibitem{debbqft}
D.~Bohm, B.~J. Hiley, and P.~N. Kaloyerou.
\newblock {An ontological basis for the quantum theory}.
\newblock {\em Phys. Rep}, 144(6):321--375, 1987.

\bibitem{randi}
S.~Alexander, L.~Freidel, and G.~Herczeg.
\newblock {An Inner Product for 4D Quantum Gravity and the Chern-Simons-Kodama
  State}.
\newblock {\em arXiv:2212.07446}, 2022.

\bibitem{durrsd}
D.~D{\"u}rr, S.~Goldstein, and N.~Zangh{\`\i}.
\newblock {Quantum motion on shape space and the gauge dependent emergence of
  dynamics and probability in absolute space and time}.
\newblock {\em J. Stat. Phys}, 180:92--134, 2020.

\bibitem{chori}
A.~Valentini.
\newblock {Quantum gravity and quantum probability}.
\newblock {\em arXiv:2104.07966}, 2021.

\bibitem{nick}
N.~G. Underwood and A.~Valentini.
\newblock {Anomalous spectral lines and relic quantum nonequilibrium}.
\newblock {\em {arXiv:1609.04576}}, {2016}.

\bibitem{d96eusch}
D.~Deutsch.
\newblock {Comment on lockwood}.
\newblock {\em {Br. J. Philos. Sci}}, 47, 1996.

\bibitem{zeh99}
H.~D. Zeh.
\newblock {Why Bohm's quantum theory?}
\newblock {\em {Found. Phys}}, 12(2), 1999.

\bibitem{brownwal}
H.~R. Brown and D.~Wallace.
\newblock {Solving the measurement problem: De Broglie--Bohm loses out to
  Everett}.
\newblock {\em {Found. Phys}}, 35(4), 2005.

\bibitem{harrikens}
N.~Harrigan and R.~W. Spekkens.
\newblock {Einstein, incompleteness, and the epistemic view of quantum states}.
\newblock {\em {Found. Phys.}}, {40}({2}), {2010}.

\bibitem{pbr}
M.~F. Pusey, J.~Barrett, and T.~Rudolph.
\newblock {On the reality of the quantum state}.
\newblock {\em {Nat. Phys.}}, {8}:{475--478}, {2012}.

\bibitem{leifer}
M.~S. Leifer.
\newblock {Is the quantum state real? An extended review of $\psi $-ontology
  theorems}.
\newblock {\em {arXiv:1409.1570}}, {2014}.

\bibitem{nist}
{\it NIST Digital Library of Mathematical Functions}.
\newblock \url{https://dlmf.nist.gov/}, Release 1.1.10 of 2023-06-15.
\newblock F.~W.~J. Olver, A.~B. {Olde Daalhuis}, D.~W. Lozier, B.~I. Schneider,
  R.~F. Boisvert, C.~W. Clark, B.~R. Miller, B.~V. Saunders, H.~S. Cohl, and
  M.~A. McClain, eds.

\end{thebibliography}

\appendix 

\section{Action of ladder operators on non-normalizable eigenstates} \label{ladder}
In orthodox quantum mechanics, lowering and raising operators provide an alternative method to obtain the quantized energy levels and eigenstates of the harmonic oscillator. It is useful to discuss the action of these operators on the non-normalizable harmonic oscillator eigenstates. We recall that the operators are defined as
\begin{align}
    \hat{a} &= \frac{1}{\sqrt{2}}(\frac{d}{dy} + y) \label{lower}\\
    \hat{a}^{\dagger} &= \frac{1}{\sqrt{2}}(-\frac{d}{dy} + y) \label{uppercut}
\end{align}
where $\hat{a}$ ($\hat{a}^{\dagger}$) is the lowering (raising) operator. It is convenient here to work with the closed form expressions of $\varphi_0^K(y)$ and $\varphi_0^K(y)$ in terms of confluent hypergeometric functions of the first kind.

\subsection{Action of lowering operator}
Let us first consider the action of $\hat{a}$ on $\varphi_0^K(y)$:
\begin{align}
    \hat{a}\varphi_0^K(y) &= e^{-y^2/2}\frac{dh_0^K(y)}{dy}\\
    &=  e^{-y^2/2}\frac{d M(\frac{1-K}{4}, \frac{1}{2} ,y^2)}{dy}\\
    &= e^{-y^2/2} (1-K)y M(\frac{3-(K-2)}{4}, \frac{3}{2}, y^2) \label{last}\\
    &= (1-K)\varphi_1^{K-2}(y) \label{wow1}
\end{align}
where, in equation (\ref{last}), we have used the identity
\begin{align}
\frac{d}{dx}M(a,b,x) = \frac{a}{b}M(a+1, b+1, x)
\end{align}
with $x=y^2$, $a=(1-K)/4$ and $b=1/2$.\\

Similarly, we can evaluate the action of $\hat{a}$ on $\varphi_1^K(y)$ as follows:
\begin{align}
    \hat{a}\varphi_1^K(y) &= e^{-y^2/2}\frac{dh_1^K(y)}{dy}\\
    &=  e^{-y^2/2}\frac{d \big(yM(\frac{3-K}{4}, \frac{3}{2} ,y^2)\big )}{dy}\\
    &= e^{-y^2/2} M(\frac{1-(K-2)}{4}, \frac{1}{2}, y^2) \label{last2}\\
    &= \varphi_0^{K-2}(y) \label{nmuch}
    \end{align}
where, in equation (\ref{last2}), we have used the identity
\begin{align}
    \frac{d}{dx}\big (x^{b-1} M(a,b,x)\big) = (b-1)x^{b-2}M(a, b-1, x)
\end{align}
  with $x=y^2$, $a=(3-K)/4$ and $b=3/2$.\\
  
Equation (\ref{wow1}) confirms that the ground state in orthodox quantum mechanics $\varphi_0^{K=1}(y)$ is annihilated by $\hat{a}$. However, it is interesting to note that $\varphi_0^{K=-1}(y)$ and $\varphi_1^{K=-1}(y)$ exist nevertheless -- they are just non normalizable. In fact, equation (\ref{nmuch}) implies that $\varphi_1^{K=1}(y)$ is transformed to $\varphi_0^{K=-1}(y)$ as a result of the action of $\hat{a}$.\\
    
Using (\ref{wow1}) and (\ref{nmuch}), we have
    \begin{align}
    \hat{a} \psi^K_{\theta, \phi}(y) &= \cos{\theta} \hat{a} \varphi_0^K(y)  + \sin{\theta} e^{i\phi} \hat{a} \varphi_1^K(y)\\
    &= \cos{\theta} (1-K) \varphi_1^{K-2}(y)  + \sin{\theta} e^{i\phi} \varphi_0^{K-2}(y)\\
    &= \cos{\theta_{K-}}\varphi_0^{K-2}(y) + \sin{\theta_{K-}} e^{i(2\pi - \phi)} \varphi_1^{K-2}(y)  = \psi^{K-2}_{\theta_{K-}, 2\pi - \phi}(y) \label{rotate1}
\end{align}
where $\cos{\theta_{K-}} = \sin{\theta}/\sqrt{\cos^2\theta(1-K)^2 + \sin^2 \theta}$ and $\sin{\theta_{K-}} = \cos{\theta}(1-K)/\sqrt{\cos^2\theta(1-K)^2 + \sin^2 \theta}$. Equation (\ref{rotate1}) implies that the action of $\hat{a}$ on $\psi^K_{\theta, \phi}(y)$ is to generate $\psi^{K-2}_{\theta_{K-}, 2\pi - \phi}(y)$. \\

\subsection{Action of raising operator}
Let us consider the action of $\hat{a}^\dagger$ on $\varphi_0^K$.
\begin{align}
\hat{a}^{\dagger}\varphi_0^K(y) &= e^{-y^2/2}(2yh_0^K(y) - \frac{dh_0^K(y)}{dy})\\
    &=  e^{-y^2/2}(2yM(\frac{1-K}{4}, \frac{1}{2} ,y^2) - \frac{d M(\frac{1-K}{4}, \frac{1}{2} ,y^2)}{dy})\\
    &= e^{-y^2/2} (1+K)y M(\frac{3-(K+2)}{4}, \frac{3}{2}, y^2) \label{last3}\\
    &= (1+K)\varphi_1^{K+2}(y) \label{wow2}
\end{align}
where, in equation (\ref{last3}), we have used the identity
\begin{align}
    \frac{d}{dx}\big (e^{-x} M(a,b,x)\big) = \frac{a-b}{b} e^{-x}M(a, b+1, x)
\end{align}
with $x=y^2$, $a=(1-K)/4$ and $b=1/2$.\\

Similarly, we can evaluate the action of $\hat{a}^\dagger$ on $\varphi_1^K$ as follows:
\begin{align}
\hat{a}^{\dagger}\varphi_1^K(y) &= e^{-y^2/2}(2yh_1^K(y) - \frac{dh_1^K(y)}{dy})\\
    &=  e^{-y^2/2}(2y^2M(\frac{3-K}{4}, \frac{3}{2} ,y^2) - \frac{d \big(yM(\frac{3-K}{4}, \frac{3}{2} ,y^2)\big )}{dy})\\
    &= -e^{-y^2/2} M(\frac{1-(K+2)}{4}, \frac{1}{2}, y^2) \label{last4}\\
    &= -\varphi_0^{K+2}(y) \label{nmuch2}
\end{align}
where, in equation (\ref{last4}), we have used the identity
\begin{align}
    \frac{d}{dx}\big (e^{-x} x^{b-1}M(a,b,x)\big) = (b-1) x^{b-2}M(a-1, b-1, x)
\end{align}
with $x=y^2$, $a=(3-K)/4$ and $b=3/2$.\\

 Note that equation (\ref{wow2}) implies that $\varphi_0^{K=-1}$ is annihilated upon being acted by $\hat{a}^\dagger$. This is similar to the annihilation of the ground state $\varphi_0^{K=+1}$ upon being acted by $\hat{a}$. \\

Using (\ref{wow2}) and (\ref{nmuch2}), we have
    \begin{align}
    \hat{a}^{\dagger} \psi^K_{\theta, \phi}(y) &= \cos{\theta} \hat{a}^{\dagger} \varphi_0^K(y)  + \sin{\theta} e^{i\phi} \hat{a}^{\dagger} \varphi_1^K(y)\\
    &= \cos{\theta} (1+K) \varphi_1^{K+2}(y)  + \sin{\theta} e^{i(\phi+\pi)} \varphi_0^{K+2}(y)\\
    &= \cos{\theta_{K+}}\varphi_0^{K+2}(y) + \sin{\theta_{K+}} e^{i(\pi -\phi)} \varphi_1^{K+2}(y) = \psi^{K+2}_{\theta_{K+}, \pi - \phi}(y)\label{rotate2}
\end{align}
where $\cos{\theta_{K+}} = \sin{\theta}/\sqrt{\cos^2\theta(1+K)^2 + \sin^2 \theta}$, $\sin{\theta_{K+}} = \cos{\theta}(1+K)/\sqrt{\cos^2\theta(1+K)^2 + \sin^2 \theta}$. Similar to equation (\ref{rotate1}), we see from (\ref{rotate2}) that the action of $\hat{a}^\dagger$ on $\psi^K_{\theta, \phi}(y)$ is to generate $\psi^{K+2}_{\theta_{K+}, \pi - \phi}(y)$. \\

From equations (\ref{wow1}), (\ref{nmuch}), (\ref{wow2}) and (\ref{nmuch2}), we see that $\hat{a}$ and $\hat{a}^\dagger$ continue to act as lowering and raising operators respectively in the general scenario where normalizability is violated. However, unlike in the normalizable scenario, it is not possible to obtain all the quantum states by successively using $\hat{a}$ and $\hat{a}^\dagger$ on a single eigenstate corresponding to a single $K$. This is because, firstly, $K$ can take values from $(-\infty, +\infty)$ once normalizability is dropped. Secondly, equations (\ref{wow1}) and (\ref{wow2}) imply that $\hat{a}\varphi_0^{K=1}$ and $\hat{a}^\dagger\varphi_0^{K=-1}$ are zero. Thirdly, the eigenstates are doubly degenerate at any value of $K$. Therefore, although both the analytic method and ladder operator approach are equivalent in orthodox quantum mechanics, the analytic method turns out to be more general when the normalizability assumption is dropped. \\

\section{Approximate form of the eigenstate at large $\pm y$}
Consider a harmonic-oscillator eigenstate $\psi^K(y)$. We know from the main text that 
\begin{align}
\psi^K(y) = e^{-y^2/2}\bigg[ a_0 M(\frac{1}{4}(1-K), \frac{1}{2}, y^2)  + a_1 y M(\frac{1}{4}(3-K), \frac{3}{2}, y^2)\bigg ]
\end{align}
Using the asymptotic form \cite{nist} of $M(c,d,y)$ as $y \to \infty$
\begin{align}
    M(c,d,y) \sim \frac{e^y y^{c-d}}{\Gamma(c)} \sum_{s=0}^\infty \frac{(1-c)_s(d-c)_s}{s!} y^{-s}
\end{align}
valid for $c \neq 0,-1, -2...$, we find that
\begin{align}
    M(\frac{1}{4}(1-K), \frac{1}{2}, y^2) &= \frac{e^{y^2}y^{-\frac{1+K}{2}}}{\Gamma(\frac{1-K}{4})}\big \{1 + \frac{(3+K)(1+K)}{16y^2}\big \} \label{a1} \\
    y M(\frac{1}{4}(3-K), \frac{3}{2}, y^2) &= \frac{e^{y^2}y^{-\frac{1+K}{2}}}{\Gamma(\frac{3-K}{4})}\big \{1 + \frac{(3+K)(1+K)}{16y^2}\big \} \label{a2}
\end{align}
where we have retained terms only up to $1/y^2$. Note that $K = 1 + 4n$ ($K = 3 + 4n$) for $M(\frac{1}{4}(1-K), \frac{1}{2}, y^2)$ $\big ( M(\frac{1}{4}(3-K), \frac{3}{2}, y^2) \big )$ if $c = 0, -1, -2...$ in $M(c,d,y)$. For these values of $K$, $M(c(K), d, y)$ is a polynomial of finite order as the power series terminates, and $e^{-y^2/2}M(c, d, y)$ vanishes at $|y| \to \infty$. Therefore, we need only concern ourselves with the non-normalizable part of $\psi^K(y)$ to determine its asymptotic behaviour. Using (\ref{a1}), (\ref{a2}), the asymptotic form of the quantum state can be written as
\begin{align}
    \lim_{|y| \to \infty} \psi^K(y) = e^{\frac{y^2}{2}}y^{-\frac{1+K}{2}}\big \{1 + \frac{(3+K)(1+K)}{16y^2} \}
\end{align}
where have eliminated the global phase and magnitude.

\end{document}